\documentclass[12pt,draftcls,onecolumn]{IEEEtran}
\usepackage{graphicx}
\usepackage{amstext}
\usepackage{algorithm}
\usepackage{algorithmic}
\usepackage{mathrsfs}
\usepackage{amssymb}
\usepackage{amsmath}
\usepackage{bm}
\allowdisplaybreaks[4]
\usepackage{epstopdf}
\usepackage{multicol}
\usepackage{stfloats}
\usepackage{url}
\usepackage{color}
\usepackage{enumerate}
\usepackage{breqn}
\usepackage{bbm}
\usepackage{cite}
\usepackage{subfig}
\usepackage{caption}
\usepackage{enumitem}

\newtheorem{Theorem}{Theorem}
\newtheorem{Lemma}{Lemma}

\newtheorem{Problem}{Problem}
\begin{document}
\title{Enhancing Performance of Random Caching in Large-Scale Heterogeneous Wireless Networks with Random Discontinuous Transmission}
\author{\IEEEauthorblockN{Wanli Wen, Ying Cui, Fu-Chun Zheng, Shi Jin and Yanxiang Jiang}\thanks{Y. Cui is with the Department of Electronic Engineering, Shanghai Jiao Tong University, China. W. Wen, F.-C. Zheng, S. Jin and Y. Jiang are with the National Mobile Communications Research Laboratory, Southeast University, Nanjing, China. This paper will be presented in part at IEEE WCNC 2018.}}

\maketitle
\vspace{-1cm}
\begin{abstract}
To make better use of file diversity provided by random caching {and improve the successful transmission probability (STP) of a file}, we consider retransmissions with {random discontinuous transmission (DTX) in a large-scale {cache-enabled} heterogeneous wireless network (HetNet) employing random caching.}  {We analyze and optimize the STP} in two mobility scenarios, {i.e.,} the high mobility scenario and the static scenario.
 First, in each scenario, by using tools from stochastic geometry, we  {obtain a closed-form expression for} the STP in the general  signal-to-interference ratio (SIR) threshold regime.  {The analysis shows that a larger caching probability corresponds to a higher STP in both scenarios;}  {random DTX can improve the STP in the static scenario and its benefit gradually diminishes when mobility increases}. In each scenario, we also derive a closed-form expression for the asymptotic outage probability in the low SIR threshold regime. The asymptotic analysis shows that the diversity gain is jointly affected by random caching and random DTX in both scenarios.
 Then, { in each scenario,} we {consider the maximization of} the STP {with respect to} the caching probability and the BS activity probability, which is a challenging {non-convex} optimization problem. {In particular, i}n the high mobility scenario, we obtain a globally optimal solution   using interior point method. In the static scenario, we develop a low-complexity iterative algorithm to obtain  a stationary point {using alternating optimization}.
 Finally, numerical results show that the proposed solutions achieve significant gains over existing baseline schemes and can well adapt to the changes of the system parameters to wisely utilize storage resources {and transmission opportunities}.
\end{abstract}
%

\IEEEpeerreviewmaketitle

\section{Introduction}

\IEEEPARstart{W}{ith} the proliferation of  smart mobile devices and multimedia services, the global mobile data traffic is expected to increase exponentially in the coming years.  However, the majority of such traffic is asynchronously but repeatedly requested by many users at different times and thus a tremendous amount of mobile data traffic have actually been redundantly generated over networks \cite{Cachingattheirelessedgedesignaspectschallengesfuturedirections}. Motivated by this, caching at base stations (BSs) has been proposed as
a promising approach {for reducing delay and backhaul load \cite{WirelesscachingMag}}.  When the coverage regions of different BSs overlap, a user can fetch the desired file from multiple adjacent BSs, and hence the performance can be increased by caching different files among BSs, i.e., providing file diversity. {In \cite{Cache-enabledsmallcellnetworksmodelingandtradeoffs, CacheenabledheterogeneouscellularnetworksComparisonandtradeoffs, Tamoorulhassan2015Modeling, ALearningBasedApproachtoCachinginHeterogenousSmallCellNetworks, OptimalGeographicCachingInCellularNetworks, CacheEnabledHeterogeneousCellularNetworksOptimalTierLevelContentPlacement, OptimalContentPlacementOffloadingHetNets, OptimizationandAnalysisofProbabilisticCachinginNtierHeterogeneousNetworks, 2017arXiv170607903W, CuiCacheWirelessSignleTier, AnalysisandOptimizationofCachingandMulticastinginLargeScaleCacheEnabledHeterogeneousWirelessNetworks, RandomCachingBasedCooperativeTransmissioninHeterogeneousWirelessNetworks, D2Dcontentdeliverynetworking2014,2018arXiv180102743J}, the authors analyze the performance of various caching designs in large-scale cache-enabled wireless networks. In particular, in} \cite{Cache-enabledsmallcellnetworksmodelingandtradeoffs} and \cite{CacheenabledheterogeneouscellularnetworksComparisonandtradeoffs}, the authors study the most popular caching design, where each BS only stores the most popular files. {As} the most popular caching {design} can not provide any spatial file diversity{, it may not} yield the optimal network performance. To provide more spatial file diversity, the authors in \cite{Tamoorulhassan2015Modeling, ALearningBasedApproachtoCachinginHeterogenousSmallCellNetworks, OptimalGeographicCachingInCellularNetworks,CacheEnabledHeterogeneousCellularNetworksOptimalTierLevelContentPlacement, OptimalContentPlacementOffloadingHetNets, OptimizationandAnalysisofProbabilisticCachinginNtierHeterogeneousNetworks, 2017arXiv170607903W, CuiCacheWirelessSignleTier,  AnalysisandOptimizationofCachingandMulticastinginLargeScaleCacheEnabledHeterogeneousWirelessNetworks, RandomCachingBasedCooperativeTransmissioninHeterogeneousWirelessNetworks, D2Dcontentdeliverynetworking2014, 2018arXiv180102743J} {consider} random caching. Specifically, in \cite{Tamoorulhassan2015Modeling}, the authors consider uniform caching where each BS randomly stores a file according to the uniform distribution. In \cite{ALearningBasedApproachtoCachinginHeterogenousSmallCellNetworks}, the authors consider i.i.d. caching where each BS stores a file in an i.i.d. manner. In \cite{OptimalGeographicCachingInCellularNetworks, CacheEnabledHeterogeneousCellularNetworksOptimalTierLevelContentPlacement, OptimalContentPlacementOffloadingHetNets, OptimizationandAnalysisofProbabilisticCachinginNtierHeterogeneousNetworks,  2017arXiv170607903W, CuiCacheWirelessSignleTier, AnalysisandOptimizationofCachingandMulticastinginLargeScaleCacheEnabledHeterogeneousWirelessNetworks, RandomCachingBasedCooperativeTransmissioninHeterogeneousWirelessNetworks,  D2Dcontentdeliverynetworking2014, 2018arXiv180102743J}, {besides analysis, the authors also consider the optimization of} random caching to either maximize the cache hit probability \cite{OptimalGeographicCachingInCellularNetworks, CacheEnabledHeterogeneousCellularNetworksOptimalTierLevelContentPlacement}, the successful offloading probability \cite{OptimalContentPlacementOffloadingHetNets} and the successful transmission probability {(STP)} \cite{OptimizationandAnalysisofProbabilisticCachinginNtierHeterogeneousNetworks,  2017arXiv170607903W, CuiCacheWirelessSignleTier, AnalysisandOptimizationofCachingandMulticastinginLargeScaleCacheEnabledHeterogeneousWirelessNetworks, RandomCachingBasedCooperativeTransmissioninHeterogeneousWirelessNetworks, 2018arXiv180102743J} or minimize {the} average caching failure probability \cite{D2Dcontentdeliverynetworking2014}. {Note that, in \cite{Tamoorulhassan2015Modeling, ALearningBasedApproachtoCachinginHeterogenousSmallCellNetworks, OptimalGeographicCachingInCellularNetworks, CacheEnabledHeterogeneousCellularNetworksOptimalTierLevelContentPlacement, OptimalContentPlacementOffloadingHetNets, OptimizationandAnalysisofProbabilisticCachinginNtierHeterogeneousNetworks,  2017arXiv170607903W, CuiCacheWirelessSignleTier, AnalysisandOptimizationofCachingandMulticastinginLargeScaleCacheEnabledHeterogeneousWirelessNetworks, RandomCachingBasedCooperativeTransmissioninHeterogeneousWirelessNetworks, D2Dcontentdeliverynetworking2014, 2018arXiv180102743J}, a user may associate
with a relatively farther BS when nearer BS{s} do not cache the requested file \cite{2018arXiv180102743J}. In this case,
the signal is usually weak compared with the interference, and the user may not successfully receive the requested file and benefit from {the} file diversity offered by random caching.}

{Increasing the number of transmissions of a file can increase the probability that eventually the file is successfully transmitted, at the cost of delay increase. Enabling retransmissions at BSs is an effectively way to improve the {STP} for}  some applications without strict delay requirements, {e.g., elastic services}. The interferences experienced by a user at different slots are usually correlated as they come from the same set of BSs \cite{STofInterferenceinAloha2009}.  In \cite{STofInterferenceinAloha2009, DPolynomialsHaenggi, localdelayinPosNetHaenggi, ManagingInterferenceCorrelationZhongYi,  HetNetsWithRandomDTXSchemeLocalDelayandEnergyEfficiency}, the authors {study retransmissions and} show that the interference correlation significantly {degrades} the performance, e.g., the diversity gain \cite{DPolynomialsHaenggi} or the transmission delay \cite{localdelayinPosNetHaenggi, ManagingInterferenceCorrelationZhongYi,  HetNetsWithRandomDTXSchemeLocalDelayandEnergyEfficiency}. In {\cite{STofInterferenceinAloha2009, DPolynomialsHaenggi, localdelayinPosNetHaenggi, ManagingInterferenceCorrelationZhongYi,  HetNetsWithRandomDTXSchemeLocalDelayandEnergyEfficiency}, the authors adopt} random discontinuous transmission (DTX) {together with retransmissions to} effectively manage such {interference} correlation, {by creating randomness for interferers, and analyze the performance in large-scale wireless networks.}   Note that, \cite{STofInterferenceinAloha2009, DPolynomialsHaenggi, localdelayinPosNetHaenggi, ManagingInterferenceCorrelationZhongYi,  HetNetsWithRandomDTXSchemeLocalDelayandEnergyEfficiency} do not consider caching at BSs.

Recently, \cite{EffectofRetransmissionsonOptimalCachinginCacheenabledSmallCellNetworks} studies the effect of retransmissions on the {performance of} random caching, and analyze{s} and optimize{{s} the {STP} in a {large-scale} single tier network.  {Note that \cite{EffectofRetransmissionsonOptimalCachinginCacheenabledSmallCellNetworks} does not consider DTX, and hence the gain of retransmissions is limited due to the strong interference correlation across multiple retransmissions. Therefore, it is still not clear how retransmissions with random DTX can maximally improve the performance of random caching. Heterogeneous wireless networks (HetNets) can further improve the network capacity by deploying small BSs {together with traditional macro BSs, to provide better time or frequency reuse}. Caching at small BSs can effectively  alleviate the backhaul  capacity {requirement} in HetNets. {For} cache-enabled HetNets, it is {also} not known how to jointly design random caching and random DTX across different {tiers}.}

In this paper, we would like to address the above {issues}. {We consider a large-scale cache-enabled HetNet. We adopt random caching and a simple retransmission protocol with random DTX to improve  the STP,  which is defined as the probability that a file can be successfully transmitted to a user. Our focus is on the analysis and optimization of joint random caching and random DTX in two  scenarios of user mobility, i.e., the high mobility scenario and the static scenario.}   The main contributions of the paper are summarized below.
\begin{itemize}
  \item First, we analyze the STP in both   scenarios. The random caching and retransmission with random DTX make the analysis very challenging. {In each scenario, b}y carefully considering the joint impacts of random caching and random DTX on the distribution of the signal-to-interference ratio (SIR) in each slot, we derive closed-form expression for the STP in the general SIR threshold regime, utilizing tools from stochastic geometry. {The analysis shows} that a larger caching probability corresponds {to} a higher STP in both scenarios, which reveals the advantage of caching. In addition, {the analysis reveals} that {random DTX can improve the STP in the static scenario and its benefit gradually diminishes when mobility increases}. We also derive a closed-form expression for the asymptotic outage probability in the low SIR threshold regime, utilizing series expansion of some special functions{. The asymptotic analysis shows} that the diversity gain is {jointly} affected by random caching and random DTX.
  \item Next,  we consider the maximization of the STP with respect to the caching probability and the BS activity probability {in both scenarios}, which is a challenging non-convex optimization problem. In the high mobility scenario,   we obtain a globally optimal solution using interior point method. In the static scenario,  we develop a low-complexity {iterative} algorithm to obtain a stationary point {using alternating optimization}.
  \item  Finally, numerical results show that the proposed solutions achieve significant gains over existing baseline schemes and can well adapt to the changes of the system parameters to wisely utilize storage resources {and transmission opportunities}. {As the maximum number of transmissions increases, more files are stored and more BSs are silenced for {improving} the STP.}
\end{itemize}

\section{System Model}
\subsection{Network Model}
We consider a large-scale HetNet consisting of $K$ independent network tiers. We denote the set of $K$ tiers by $\mathcal{K}=\{1,2,\cdots,K\}$. All network tiers are co-channel deployed. The BS locations in tier $k$ are modeled by an independent homogeneous Poisson point process (PPP) $\Phi_k$ with density $\lambda_k$. {Let $\Phi$ be the superposition of $\Phi_k$, $k\in\mathcal{K}$, i.e., $\Phi\triangleq\bigcup_{k\in\mathcal{K}}\Phi_k$, which denotes the locations of all tiers of BSs in the network.} Each BS in tier $k$ has one transmit antenna with transmission power $P_k$. For the propagation model, we consider a general power-law path-loss model in which a transmitted signal from a BS with distance $r$, is attenuated by a factor $r^{-\alpha}$, where $\alpha>2$ denotes the path-loss exponent. For the small-scale fading model, we assume Rayleigh fading. Since a HetNet is primarily interference-limited, we ignore the thermal noise for simplicity.

Each user has one receive antenna. We consider a discrete-time system with time being slotted. Let $t=1,2,\cdots$ denote the slot index. Two {scenarios} of user mobility are considered: the high mobility scenario and the static scenario \cite{localdelayinPosNetHaenggi}. In the high mobility scenario, in each slot, the user locations follow an independent homogeneous PPP, i.e., a new realization of the PPP for the user locations is drawn in each slot, and the user locations are independent over time. Mathematically, the high mobility scenario is equivalent to the case {where} the user locations follow an independent homogeneous PPP and stay fixed over time, and in each slot, new realizations of the $K$ independent PPPs for the $K$ tiers of BSs are drawn {\cite{localdelayinPosNetHaenggi}}. In the static scenario, the user locations stay fixed over time and follow an independent PPP. {Note that, the mobility scenario  in a practical network  is between the two scenarios, and hence the results in this paper provide some theoretical performance bounds for a practical network.} In both scenarios, without loss of generality (w.l.o.g.), we can study the performance of a typical user $u_0$, which is located at the origin $o$, according to Slivnyak's Theorem {\cite{Haenggi2012Stochastic}}.

Let ${\mathcal{N}} = \{ 1,2, \cdots ,N\} $ denote the set of $N$ files in the HetNet. For ease of analysis, as in \cite{
OptimizationandAnalysisofProbabilisticCachinginNtierHeterogeneousNetworks,  AnalysisandOptimizationofCachingandMulticastinginLargeScaleCacheEnabledHeterogeneousWirelessNetworks}, we assume that all files have the same size, and file popularity distribution is identical among all users.\footnote{Note that, the results in this paper can be easily extended to the case of different file sizes. To be specific, we can consider file combinations of the same total size, but formed by files of possibly different sizes.} The probability that file $n\in\mathcal{N}$ is requested by each user is $a_n\in(0,1)$, where $\sum\nolimits_{n\in\mathcal{N}} {{a_n}}  = 1$. Thus, the file popularity distribution is given by $\mathbf{a}\triangleq (a_n)_{n\in\mathcal{N}}$, which is assumed to be known a priori.\footnote{Note that, the file popularity evolves at a slower timescale and various learning methodologies can be employed to estimate the file popularity over time \cite{FemtoCachingWirelessContentDeliveryThroughDistributedCachingHelpers}.}  In addition, w.l.o.g., we assume that ${a_1} \ge {a_2} \ge  \cdots \ge {a_N}$, i.e., the popularity rank of file $n$ is $n$. Assume that at the beginning of slot 1, each user randomly requests a file according to the file popularity distribution $\mathbf{a}$. We shall consider the delivery of each requested file over $M$ consecutive slots. The network consists of cache-enabled BSs. In particular, each BS in tier $k$ is equipped with a cache of size $C_k\le N$ to store $C_k$ different files out of $N$. \vspace{-3mm}

\subsection{Random Caching {and Retransmissions with Random DTX}}
To provide high spatial file diversity, we consider a random caching design similar to the one in \cite{CuiCacheWirelessSignleTier}, where file $n$ is stored at each BS in tier $k$ according to a certain probability $T_{n,k}\in[0,1]$, called the caching probability of file $n$ in tier $k$. Denote $\mathbf{T}\triangleq \left(\mathbf{T}_n\right)_{n\in\mathcal{N}}\in[0,1]^{NK\times 1}$, where $\mathbf{T}_n\triangleq(T_{n,k})_{k
\in\mathcal{K}}\in[0,1]^{K\times 1}$, as the caching distribution of the $N$ files in the $K$-tier HetNet. Note that, the random caching design is parameterized by $\mathbf{T}$. We have  \cite{AnalysisandOptimizationofCachingandMulticastinginLargeScaleCacheEnabledHeterogeneousWirelessNetworks, CuiCacheWirelessSignleTier}:\footnote{To implement the random caching design, we randomly place a file combination of $C_k$ different files at each BS in tier $k$ according to a corresponding caching probability for file combinations. The detailed relationship between $\mathbf{T}$ and the caching probability for file combinations can be found in \cite{CuiCacheWirelessSignleTier}.}
\begin{eqnarray}\label{eqsysmod1}
&&0\le T_{n,k} \le 1, n\in\mathcal{N}, k\in\mathcal{K},\label{eqconstcachingprob}\\
&&\sum_{n\in \mathcal{N}}T_{n,k}=C_k, k\in\mathcal{K}.\label{eqconst2mbcachesize}
\end{eqnarray} Let $\Phi_{n,k}$ denote the point process of the BSs in tier $k$ which store file $n$. Note that, $\Phi_{n,k}\subseteq\Phi_k$, $n\in\mathcal{N}$. Under the random caching design, $\Phi_{n,k}$, $n\in\mathcal{N}$ are independent PPPs with densities $\lambda_kT_{n,k}$, $n\in\mathcal{N}$.

Consider a user requesting file $n$ at the beginning of slot 1. If file $n$ is not stored in any tier, the user will not be served. Otherwise, the user is associated with the BS which not only stores file $n$ but also provides the maximum  average {received signal strength (RSS)} \cite{CacheEnabledHeterogeneousCellularNetworksOptimalTierLevelContentPlacement} among all BSs in the $K$-tier HetNet, referred to as its serving BS. Note that, in the high mobility scenario, the user association changes from slot to slot{, while} in the static scenario, the user association does not change over slots. Under this content-{based} user association, in each slot, a user may not be associated with the BS {which provides} the maximum average RSS {if it has not stored file $n$}. As a result, a user may suffer from more severe inter-cell interference under this content-{based}  user association than under the traditional connection-based user association.

The transmission of a file in one slot {is more likely to} fail, if $u_0$ is not associated with the BS {providing the maximum  average RSS.} Increasing the number of transmissions of a file can increase the probability that eventually the file is successfully transmitted, at the cost of delay increase. In addition, there are some applications without strict delay requirements, {e.g., elastic services}. Therefore, for those applications, we consider a simple retransmission protocol in which a file is repeatedly transmitted until it is successfully received or the number of transmissions exceeds $M$.\footnote{Note that, we consider the case that a user will not request any new file until the current file request is served or expires (i.e., {the number of transmissions} exceeds $M$).}

{However, in a practical HetNet, the interference suffered by a user is temporally correlated since it comes from the same set of interferers in different time slots \cite{ManagingInterferenceCorrelationZhongYi}. Such correlation makes the SIRs temporally correlated and thus dramatically decreases the  performance gain {of} retransmission.} In order to {manage such correlation,} we consider random DTX at the BSs {\cite{HetNetsWithRandomDTXSchemeLocalDelayandEnergyEfficiency, ManagingInterferenceCorrelationZhongYi},} where each BS has two possible transmission states in each slot, i.e., the active state and the inactive state. Specifically, in each slot, a BS is active with probability $\beta\in(0,1]$, called the activity probability, and is inactive with probability $1-\beta$, independent of the BS location and slot.\footnote{Note that, a user cannot be served in a slot if its serving BS is inactive in this slot.} Note that, the random DTX design is parameterized by $\beta$. The density of active BSs in tier $k$ is $\beta \lambda_k$.

\subsection{Performance Metric}\label{subsectionPeromanceMetric}
Suppose {that the typical user} $u_0$ requests file $n$ at the beginning of slot 1. Let $k_0$ denote the index of the tier with which $u_0$ is
associated {and} $l_0\in\Phi_{k_0}$ denote the index of the serving BS of $u_0$. We denote $X_{k,l,0}(t)$ and $h_{k,l,0}(t)$ as the distance and the fading power coefficient between BS $l\in\Phi_{k}$ and $u_0$ in slot $t$, respectively. {Assume $h_{k,l,0}(t)$, $l\in\Phi$, $t=1,2,\cdots,M$ are i.i.d., according to the exponential distribution with unit mean.} Let $\mathcal{B}_k^a(t)$ be the set of active BSs in tier $k$ in slot $t$. When $u_0$ requests file $n$ and file $n$ is transmitted by BS {$l_0$}, the signal-to-interference ratio (SIR) of $u_0$ in slot $t$ is given by
{\begingroup\makeatletter\def\f@size{11}\check@mathfonts
\def\maketag@@@#1{\hbox{\m@th\normalsize\normalfont#1}}\setlength{\arraycolsep}{0.0em}\setlength{\arraycolsep}{0.0em}
\begin{equation}\label{eqdefSIR}
\mathrm{SIR}_{n,0}(t)=\frac{P_{k_0}h_{k_0,l_0,0}(t)X_{k_0,l_0,0}(t)^{-\alpha}(t)\mathbbm{1}(l_0\in\mathcal{B}_{k_0}^a(t))} {\sum\limits_{l\in\Phi_{k_0}\setminus\{l_0\}}P_{k_0}h_{k_0,l,0}(t)X_{k_0,l,0}^{-\alpha}(t) \mathbbm{1}(l\in\mathcal{B}_{k_0}^a(t))+ \sum\limits_{j\in\mathcal{K}\setminus \{k_0\}}\sum\limits_{l\in\Phi_j}P_jh_{j,l,0}(t)X_{j,l,0}^{-\alpha}(t) \mathbbm{1}(l\in\mathcal{B}_j^a(t))},
\end{equation}\setlength{\arraycolsep}{5pt}\endgroup}where $\mathbbm{1}(\cdot)$ denotes the indicator function.

We say that file $n$ is successfully transmitted to $u_0$ in slot $t$ if $\mathrm{SIR}_{n,0}(t)$ is greater than or equal to a given threshold $\theta$, i.e., $\mathrm{SIR}_{n,0}(t)\ge \theta$. Let $\mathcal{S}_n(t)$ denote the event that file $n$ is successfully transmitted to $u_0$ in slot $t$ and $\mathcal{S}^c_n(t)$ denote the complementary event of $\mathcal{S}_n(t)$, i.e., the event that file $n$ is not successfully transmitted to $u_0$ in slot $t$. The probability that file $n$ is successfully transmitted to $u_0$ in $M$ consecutive slots, referred to as the successful transmission probability (STP) of file $n$, under the adopted simple retransmission protocol {in the high mobility and static scenarios}, is given by
{\setlength{\arraycolsep}{0.0em}
\begin{eqnarray}\label{eqqni}
q_{n,i}(\mathbf{T}_n,\beta)=1-\Pr\left(\mathcal{S}^c_n(1),\mathcal{S}^c_n(2),\cdots,\mathcal{S}^c_n(M)\right),\quad i\in\{\mathrm{hm},\mathrm{st}\}.\label{eqDefqni}
\end{eqnarray}\setlength{\arraycolsep}{5pt}}\vspace{-0.8cm}

In the high mobility scenario, since the events $\mathcal{S}_n(t)$ (or $\mathcal{S}^c_n(t)$), $t=1,2,\cdots,M$, are i.i.d., the STP of file $n$ in (\ref{eqqni}) can be expressed as
{\setlength{\arraycolsep}{0.0em}
\begin{eqnarray}
q_{n,\mathrm{hm}}(\mathbf{T}_n,\beta)&=& 1-\prod_{t=1}^{M}\Pr\left(\mathcal{S}^c_n(t)\right) =1-\left(1-\Pr\left(\mathcal{S}_{n}\right)\right)^{M},\label{eqDefphihm}
\end{eqnarray}\setlength{\arraycolsep}{5pt}}where $\Pr\left(\mathcal{S}_{n}\right)=\Pr\left(\mathrm{SIR}_{n,0}(t)\ge \theta\right)$ {is the STP of file $n$ in one slot.} Here, we have dropped the index $t$ in $\mathcal{S}_n(t)$, as $\mathcal{S}_n(t)$, $t=1,2,\cdots,M$, are i.i.d.

In the static scenario, as the locations of BSs and $u_0$ do not change, the events $\mathcal{S}^c_n(t)$, $t=1,2,\cdots,M$, are correlated.  Let {$\mathcal{S}_{n|\Phi}(t)$} denote the event that file $n$ is successfully transmitted to $u_0$ in slot $t$, conditioned on $\Phi$. Similarly, $\mathcal{S}^c_{n|\Phi}(t)$ denotes the complementary event of $\mathcal{S}_{n|\Phi}(t)$. Note that, the events $\mathcal{S}_{n|\Phi}(t)$ (or $\mathcal{S}_{n|\Phi}^c(t)$), $t=1,2,\cdots,M$, are i.i.d. due to the fact that the fading power coefficients are i.i.d. with respect to (w.r.t.) $t$. Thus, in the static scenario, the STP of file $n$ in (\ref{eqqni}) can be expressed as
{\setlength{\arraycolsep}{0.0em}
\begin{eqnarray}
q_{n,\mathrm{st}}(\mathbf{T}_n,\beta)&=&\mathbbm{E}_{\Phi} \left(1-\Pr\left(\mathcal{S}^c_{n|\Phi}(1), \mathcal{S}^c_{n|\Phi}(2),\cdots,\mathcal{S}^c_{n|\Phi}(M)\right)\right)\nonumber\\
&=& \mathbbm{E}_{\Phi} \left(1-\prod_{t=1}^{M}\Pr\left(\mathcal{S}^c_{n|\Phi}(t)\right)\right)= \mathbbm{E}_{\Phi} \left(1-\left(1-\Pr\left(\mathcal{S}_{n|\Phi}\right)\right)^{M}\right),\label{eqDefphist}
\end{eqnarray}\setlength{\arraycolsep}{5pt}}where $\mathbbm{E}(\cdot)$ is the expectation operation and $\Pr\left(\mathcal{S}_{n|\Phi}\right)=\Pr\left(\mathrm{SIR}_{n,0}(t)\ge \theta|\Phi\right)$.  {Note that, $\mathbbm{E}_{\Phi}\left(\Pr\left(\mathcal{S}_{n|\Phi}\right)\right)=\Pr\left(\mathcal{S}_{n}\right)$.} Here, we have dropped the index $t$ in $\mathcal{S}_{n|\Phi}(t)$, as $\mathcal{S}_{n|\Phi}(t)$, $t=1,2,\cdots,M$, are i.i.d..

Note that, $f(x)=1-(1-x)^M$, $x\in[0,1]$, is a linear function when $M=1$, and  a concave function when $M=2,3,\cdots$. Thus, by Jensen's inequality, we have $q_{n,\mathrm{st}}(\mathbf{T}_n,\beta)\le q_{n,\mathrm{hm}}(\mathbf{T}_n,\beta)$, where the equality holds {when} $M=1$, implying that mobility has a positive effect on the STP. The intuitions are given as follows. In the static scenario, the locations of BSs during $M$ consecutive slots stay fixed, leading to temporal SIR correlation. That is, if the transmission in one slot fails, there is a higher chance that  the transmission in another slot also fails. In contrast, in the high mobility scenario, the locations of BSs during $M$ consecutive slots are independent, and hence there is no correlation among SIRs during $M$ consecutive slots. Consequently, a user has a higher chance to experience a favorable transmission channel with high SIR within $M$ transmissions. Therefore, mobility increases temporal diversity, leading to the STP increase.

Users are mostly concerned about whether their requested files can be successfully received. Therefore, in this paper, we adopt the probability that a randomly requested file by the typical user is successfully transmitted in $M$ consecutive slots, referred to as the STP, as the network performance metric. By total probability theorem, the STP {in the high mobility and static scenarios} is given by
{\setlength{\arraycolsep}{0.0em}
\begin{eqnarray}
{q}_i(\mathbf{T},\beta)&=&\sum_{n\in\mathcal{N}}a_n{q}_{n,i}(\mathbf{T}_n,\beta),\quad i\in\{\mathrm{hm},\mathrm{st}\},\label{eqDefphihmaverage}
\end{eqnarray}\setlength{\arraycolsep}{5pt}}where {$\mathbf{T}$ and $\beta$} are the design parameters of random caching and random DTX, {respectively}.

\section{High Mobility Scenario}
In this section, we consider the high mobility scenario. {We first analyze the STP and then} maximize the STP by optimizing the design parameters of random caching and random DTX.
\subsection{Performance Analysis}
In this part, we  analyze the STP in the general SIR threshold regime and the low SIR threshold regime, respectively. To be specific, we only need to analyze the STP of file $n$, i.e., $q_{n,\mathrm{hm}}(\mathbf{T}_n,\beta)$, in the two regimes. Then, by (\ref{eqDefphihmaverage}), we can directly obtain the STP, i.e., ${q}_{\mathrm{hm}}(\mathbf{T},\beta)$.
\subsubsection{Performance Analysis in General SIR Threshold Regime}
In this part, we analyze { $q_{n,\mathrm{hm}}(\mathbf{T}_n,\beta)$} in the general SIR threshold regime, using tools from stochastic geometry. To calculate $q_{n,\mathrm{hm}}(\mathbf{T}_n,\beta)$, based on (\ref{eqDefphihm}), we first need to analyze the distribution of the SIR, $\mathrm{SIR}_{n,0}(t)$. Under random caching, there are three types of interferers for $u_0$: i) all the other BSs in the same tier as the serving BS of $u_0$ which have stored the desired file of $u_0$ (apart from the serving BS of $u_0$), ii) all the BSs in the same tier as the serving BS of $u_0$ {which have not stored} the desired file of $u_0$, and iii) all the BSs in other tiers. In addition, under random DTX, the serving BS of $u_0$ is active with probability $\beta$, and the number of {interferers} of $u_0$ is $\beta$ {times} that for the case where the BSs are always active. By jointly considering the impacts of random caching and random DTX on $\mathrm{SIR}_{n,0}(t)$, we can derive the distribution of $\mathrm{SIR}_{n,0}(t)$ and then $q_{n,\mathrm{hm}}(\mathbf{T}_n,\beta)$, as summarized in the following~theorem.
\begin{Theorem}[STP in High Mobility Scenario]\label{TheoremSDPMb}
The STP of file $n$ in the high mobility scenario is given by
{\setlength{\arraycolsep}{0.0em}
\begin{eqnarray}\label{eqCorollaryphiequalximb}
q_{n,\mathrm{hm}}(\mathbf{T}_n,\beta)=1-\left(1-\frac{ \beta\sum_{k\in\mathcal{K}}z_kT_{n,k}}{W(\beta)\sum_{k\in\mathcal{K}}z_kT_{n,k} +V(\beta)(1-\sum_{k\in\mathcal{K}}z_kT_{n,k})}\right)^M,\quad n\in\mathcal{N},
\end{eqnarray}\setlength{\arraycolsep}{5pt}}where $z_k\triangleq  \frac{\lambda_kP_k^{{2}/{\alpha}}} {\sum_{j\in\mathcal{K}}\lambda_jP_j^{{2}/{\alpha}}}$, $W(\beta)\triangleq 1-\beta+\beta{_2F}_1(-\frac{2}{\alpha},1;1-\frac{2}{\alpha};-\theta)$, $V(\beta)\triangleq \beta \Gamma\left(1+\frac{2}{\alpha}\right)\Gamma\left(1-\frac{2}{\alpha}\right)\theta^{\frac{2}{\alpha}}$, and ${_2F_1}\left(a,b;c;z\right)$ and $\Gamma(x,y)$ denote the Gauss hypergeometric function and Gamma function, respectively.
\end{Theorem}

\indent\indent \textit{Proof}: See Appendix \ref{ProofTheoremSDPMb}. $\hfill\blacksquare$

Theorem \ref{TheoremSDPMb} provides a closed-form expression for {$q_{n,\mathrm{hm}}(\mathbf{T}_n,\beta)$} in the general {SIR threshold regime}. From Theorem~\ref{TheoremSDPMb}, we can see that the system parameters $K$, $M$, $\alpha$, $\lambda_k$, $k\in\mathcal{K}$, $P_k$, $k\in\mathcal{K}$, $\theta$, $\beta$ and $\mathbf{T}_n$ jointly affect $q_{n,\mathrm{hm}}(\mathbf{T}_n,\beta)$ in a complex manner.

\begin{figure*}[!t]
    \centering
        \subfloat[$T_{n,1}=0.8$ and $\beta=0.9$.]{\includegraphics[width=3.3in]{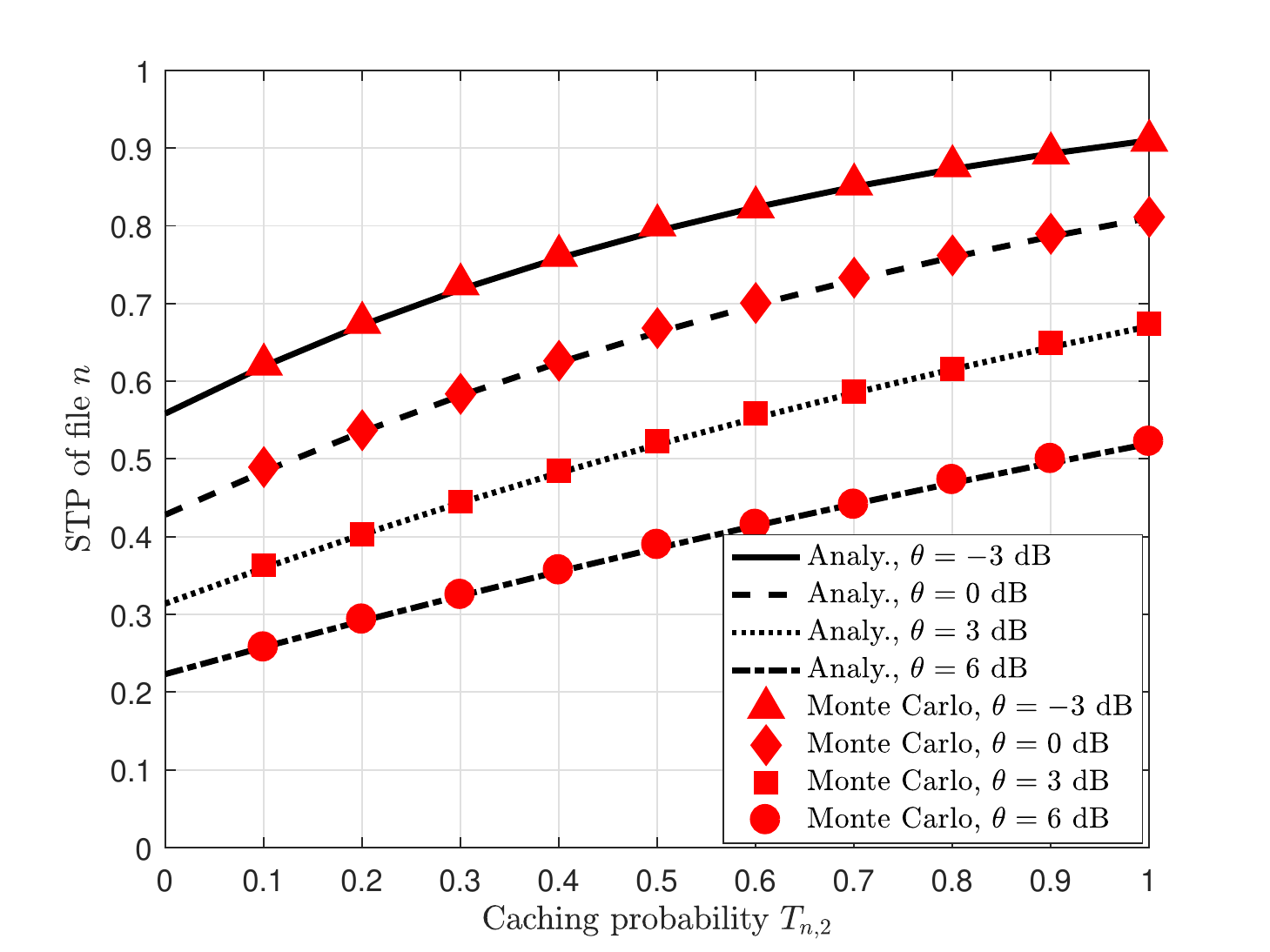}}
        \subfloat[$T_{n,1}=T_{n,2}=0.8$.]{\includegraphics[width=3.3in]{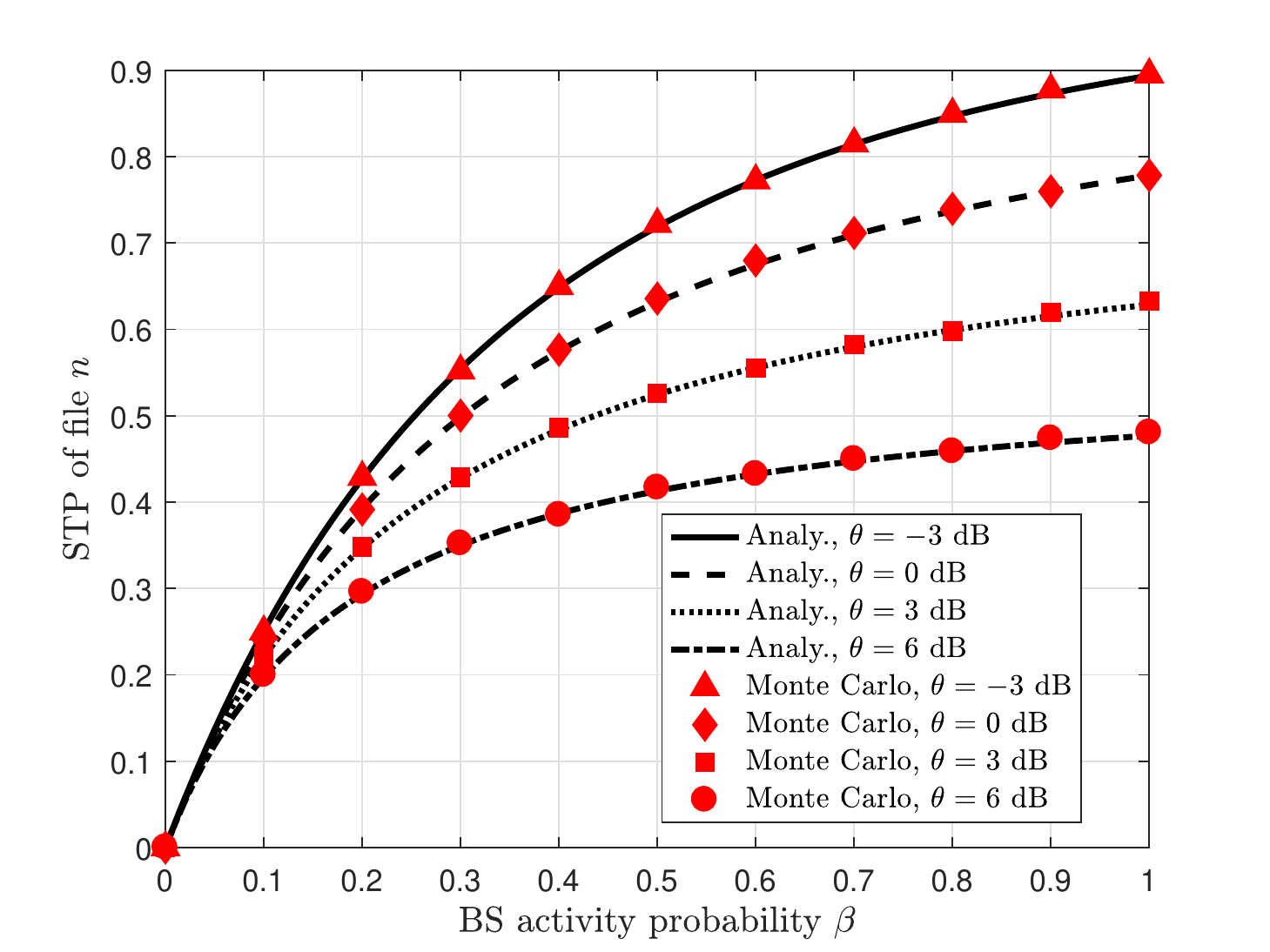}}

      \caption{{$q_{n,\mathrm{hm}}(\mathbf{T}_n,\beta)$} versus $T_{n,k}$ and $\beta$, respectively, in the high mobility scenario. $K=2$, $P_1=20$ W, $P_2=0.13$ W, $\lambda_1=\frac{1}{ 250^2\pi}$, $\lambda_2=\frac{1}{ 50^2\pi}$, $\alpha=3.5$, {and} $M=3$. The power parameters are chosen according to \cite{Howmuchenergywirelessnet2011}. In the Monte Carlo simulations, we choose a large spatial window, which is a square of $10^4 \times 10^4$ $\mathrm{m}^2$, and the final simulation results are obtained by averaging over $10^5$ independent realizations.}\label{figsdpasymptoticZero}\vspace{-0.8cm}
\end{figure*}

Based on Theorem 1, we characterize { how $q_{n,\mathrm{hm}}(\mathbf{T}_n,\beta)$ changes with $T_{n,k}$ and $\beta$}, as summarized blow.
\begin{Lemma}[Effects of Random Caching and Random DTX]\label{propMonotonicityeqCorollaryphiequalximb}
\begin{itemize}
  \item $q_{n,\mathrm{hm}}(\mathbf{T}_n,\beta)$ is an increasing and concave function of $T_{n,k}$, for all $k\in\mathcal{K}$.
  \item $q_{n,\mathrm{hm}}(\mathbf{T}_n,\beta)$ is an increasing and concave function of~$\beta$.
\end{itemize}
\end{Lemma}

The first result in Lemma \ref{propMonotonicityeqCorollaryphiequalximb} shows that a larger {$T_{n,k}$} corresponds to a larger {$q_{n,\mathrm{hm}}(\mathbf{T}_n,\beta)$}, {which reveals the advantage of caching.  This is because the average distance between a user requesting file $n$ and its serving BS decreases with $T_{n,k}$.}  The second result in Lemma \ref{propMonotonicityeqCorollaryphiequalximb} shows that a larger $\beta$ corresponds to a larger {$q_{n,\mathrm{hm}}(\mathbf{T}_n,\beta)$}, which reveals that it is not beneficial to apply random DTX in the high mobility scenario.
To understand this {result},  we first study the STP of file $n$ in {one} slot, i.e., $\Pr\left(\mathcal{S}_n\right)$. {By setting $M=1$ in (\ref{eqCorollaryphiequalximb}), we have}
{\setlength{\arraycolsep}{0.0em}
\begin{eqnarray}\label{eqSTPOneSlot}
\Pr\left(\mathcal{S}_n\right)=\frac{\beta\sum_{k\in\mathcal{K}}z_kT_{n,k}}{W(\beta)\sum_{k\in\mathcal{K}}z_kT_{n,k} +V(\beta)(1-\sum_{k\in\mathcal{K}}z_kT_{n,k})},\quad n\in\mathcal{N}.
\end{eqnarray}\setlength{\arraycolsep}{5pt}}It is easy to verify that $\Pr\left(\mathcal{S}_n\right)$ is an increasing function of $\beta$, {implying that the penalty of random DTX} in signal reduction overtakes its advantage in interference reduction in one slot. In the high mobility scenario, $\mathcal{S}_n(t)$, $t=1,2,\cdots,M$ are i.i.d., {implying that random DTX has no further benefit of reducing interference correlation. Therefore, random DTX cannot improve the STP in the high mobility scenario.}
Fig. \ref{figsdpasymptoticZero} plots {$q_{n,\mathrm{hm}}(\mathbf{T}_n,\beta)$} versus
{$T_{n,k}$} and {$\beta$}, respectively, verifying Theorem \ref{TheoremSDPMb} and Lemma~\ref{propMonotonicityeqCorollaryphiequalximb}.

\subsubsection{Performance Analysis in Low SIR Threshold Regime}

To further obtain insights, in this part, we analyze the outage probability of file $n$ which is defined as  $\bar{q}_{n,\mathrm{hm}}(\mathbf{T}_n,\beta)\triangleq 1- q_{n,\mathrm{hm}}(\mathbf{T}_n,\beta)$, in the low SIR threshold regime, i.e., $\theta\rightarrow0$, {where the (normalized) target bit rate $\tau\triangleq\log_2\left(1+\theta\right) \to 0$.\footnote{Different types of files may have different target bit rates. For instance, some video files such as MPEG 1, MPEG 4 and H.323 \cite{BluetoothandWiFiwireless} and audio files such as CD and MP3 \cite{Ganz:2003:MWN:861407} require relatively low target bit rates.}} Let $\mathbf{1}\triangleq(1,1,\cdots,1)^T\in\mathbbm{R}^K$ denote the $K$-dimensional  all-one vector. Here, $(\cdot)^T$ denotes the transpose operation. {Denote $\mathcal{T}_n\triangleq \{\mathbf{T}_n|T_{n,k}\in [0,1],k\in\mathcal{K}\}$.} {For ease of illustration, in the following, we consider four cases.
\begin{itemize}[leftmargin=1cm]
  \item \textbf{Case i)}: File $n$ is stored at each BS and random DTX is not applied, i.e., $ \mathbf{T}_n=\mathbf{1}$ and $\beta=1$.
  \item \textbf{Case ii)}: File $n$ is not stored at {any} BS and random DTX is not applied, i.e., $ \mathbf{T}_n\in\mathcal{T}_n\setminus\{\mathbf{1}\}$ and $\beta=1$.
  \item \textbf{Case iii)}: File $n$ is stored at each BS and random DTX is applied, i.e., $ \mathbf{T}_n=\mathbf{1}$ and $\beta<1$.
  \item \textbf{Case iv)}: File $n$ is not stored at {any} BS and random DTX is applied, i.e., $ \mathbf{T}_n\in\mathcal{T}_n\setminus\{\mathbf{1}\}$ and $\beta<1$.
\end{itemize}By {analyzing} the four cases,} we have the following result.\footnote{Note that, $f(x)\sim g(x)$ when $x\rightarrow 0$ means $\lim_{x\rightarrow 0}f(x)/g(x)=1.$}
\begin{Lemma}[Outage Probability in High Mobility Scenario When $\theta\to0$]\label{lemmathetaarrow0mb}
In the high mobility scenario, when $\theta\to 0$, we have
{\setlength{\arraycolsep}{0.0em}
\begin{eqnarray}\label{eqlemmaphifasyTfmb}
\bar{q}_{n,\mathrm{hm}}(\mathbf{T}_n,\beta) \sim \left(1-\beta\right)^M+c_{\mathrm{hm}}(\mathbf{T}_n,\beta),\quad n\in\mathcal{N},
\end{eqnarray}\setlength{\arraycolsep}{5pt}}where
{\begingroup\makeatletter\def\f@size{11}\check@mathfonts
\def\maketag@@@#1{\hbox{\m@th\normalsize\normalfont#1}}\setlength{\arraycolsep}{0.0em}\setlength{\arraycolsep}{0.0em}
\begin{eqnarray}\label{eqchm}
c_{\mathrm{hm}}(\mathbf{T}_n,\beta)\triangleq \begin{cases}
                            \theta^{M}\left(\frac{2}{\alpha-2}\right)^M, & \mbox{case i)} , \\
                            \theta^{\frac{2 M}{\alpha}}\left(\left(\frac{1}{\sum_{k\in\mathcal{K}}z_kT_{n,k}}-1\right) \Gamma\left(1+\frac{2}{\alpha}\right)\Gamma\left(1-\frac{2}{\alpha}\right)\right)^M, & \mbox{case ii)},  \\
                            \theta\left(1-\beta\right)^{M-1}M\beta^2\frac{2}{\alpha-2}, & \mbox{case iii)}, \\
                            \theta^{\frac{2}{\alpha}}\left(1-\beta\right)^{M-1}M\beta^2 \left(\frac{1}{\sum_{k\in\mathcal{K}}z_kT_{n,k}}-1\right)\Gamma\left(1+\frac{2}{\alpha}\right)\Gamma\left(1-\frac{2}{\alpha}\right), & \mbox{case iv)}.
                          \end{cases}
\end{eqnarray}\setlength{\arraycolsep}{5pt}\endgroup}
\end{Lemma}

\indent\indent \textit{Proof}: See Appendix \ref{Prooflemmathetaarrow0hm}. $\hfill\blacksquare$

\begin{figure*}[!t]
    \centering
        {\includegraphics[width=3.3in]{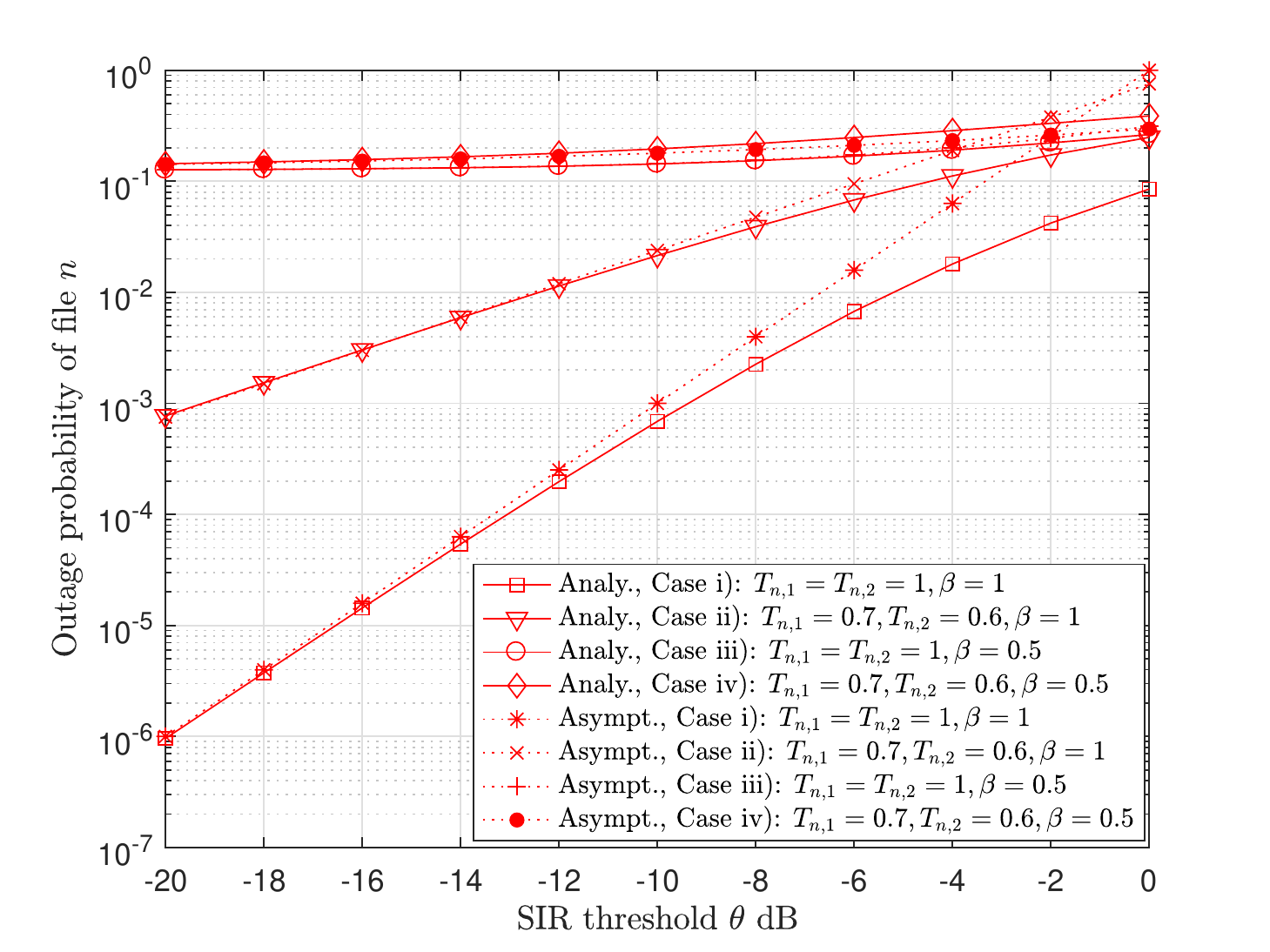}}

      \caption{{$\bar{q}_{n,\mathrm{hm}}(\mathbf{T}_n,\beta)$} versus $\theta$ in the high mobility scenario in different cases. $P_1=20$ W, $P_2=0.13$ W, $\lambda_1=\frac{1}{ 250^2\pi}$, $\lambda_2=\frac{1}{ 50^2\pi}$, $\alpha=4$, {and} $M=3$.}\label{figsdpexactandasymptosHM}\vspace{-0.8cm}
\end{figure*}

From Lemma \ref{lemmathetaarrow0mb}, we can see that both  $\mathbf{T}_n$ and  $\beta$ significantly affect the asymptotic behaviours of $\bar{q}_{n,\mathrm{hm}}(\mathbf{T}_n,\beta)$ when $\theta\to 0$, but in  different manners. Fig. \ref{figsdpexactandasymptosHM} plots {$\bar{q}_{n,\mathrm{hm}}(\mathbf{T}_n,\beta)$} versus  $\theta$ and indicates that Lemma \ref{lemmathetaarrow0mb} provides a good approximation for $\bar{q}_{n,\mathrm{hm}}(\mathbf{T}_n,\beta)$ when $\theta$ is small. {In addition, from Fig. \ref{figsdpexactandasymptosHM}, we observe that the rates of decay to zero of $\bar{q}_{n,\mathrm{hm}}(\mathbf{T}_n,\beta)$ when $\theta\to0$ in the four cases are different. In the following, we further characterize such rate {in each case}, referred to as the diversity gain \cite{AStochasticGeometryAnalysisofInterCellInterferenceCoordinationandIntraCellDiversity}, i.e.,}
{\setlength{\arraycolsep}{0.0em}
\begin{eqnarray}\label{eqdefdiversitygainhm}
d_{\mathrm{hm}}=\lim_{\theta\rightarrow0}\frac{\log(\bar{q}_{n,\mathrm{hm}}(\mathbf{T}_n,\beta))}{\log\theta}.
\end{eqnarray}\setlength{\arraycolsep}{5pt}}{Note that the definition in (\ref{eqdefdiversitygainhm}) is similar to the usual definition of diversity gain as the rate of decay to zero of the error probability in the high SNR regime \cite{DiversityinWirelessnetworkTSE}.} A larger diversity gain implies a faster decay to zero of $\bar{q}_{n,\mathrm{hm}}(\mathbf{T}_n,\beta)$ with decreasing $\theta$. {From Lemma \ref{lemmathetaarrow0mb}, we have the following result.}
\begin{Lemma}[Diversity Gain in High Mobility Scenario]\label{lemmaDiversitygainmb}
The diversity gain in the high mobility scenario is given by
{\setlength{\arraycolsep}{0.0em}
\begin{eqnarray}\label{eqdiversityorderHML1}
d_{\mathrm{hm}}=\begin{cases}
                  M, & \mbox{case i)},\\
                  \frac{2 M}{\alpha}, & \mbox{case ii)},\\
                  0, &\mbox{cases iii) and iv)}.
                \end{cases}
\end{eqnarray}\setlength{\arraycolsep}{5pt}}
\end{Lemma}

{Lemma \ref{lemmaDiversitygainmb} tells us that as long as random DTX is applied, there is no diversity gain. Without random DTX, caching a file at every BS can achieve the full diversity gain $M$, and caching  a file only at some BSs, irrespective of the caching probability, achieves the same smaller diversity gain $\frac{2M}{\alpha}\in(0,M)$ (as $\alpha>2$ and $M\ge1$). Fig. \ref{figsdpexactandasymptosHM} verifies Lemma \ref{lemmaDiversitygainmb}.}\vspace{-3mm}

\subsection{Performance Optimization}
By substituting (\ref{eqCorollaryphiequalximb}) into (\ref{eqDefphihmaverage}), the STP in the high mobility scenario is calculated as
{\setlength{\arraycolsep}{0.0em}
\begin{equation}\label{eqCorollaryphiequalxi2mb}
{q}_{\mathrm{hm}}(\mathbf{T},\beta)=\sum_{n\in\mathcal{N}}a_n \left(1-\left(1-\frac{ \beta\sum_{k\in\mathcal{K}}z_kT_{n,k}}{W(\beta)\sum_{k\in\mathcal{K}}z_kT_{n,k} +V(\beta)(1-\sum_{k\in\mathcal{K}}z_kT_{n,k})}\right)^M\right).
\end{equation}
\setlength{\arraycolsep}{5pt}}The caching distribution $\mathbf{T}$ and BS activity probability $\beta$ significantly affect the STP in the high mobility scenario. We would like to maximize ${q}_{\mathrm{hm}}(\mathbf{T},\beta)$ in (\ref{eqCorollaryphiequalxi2mb}) by jointly optimizing $\mathbf{T}$ and $\beta$. Specifically, we have the following optimization problem.
\begin{Problem}[{Optimization of Random Caching and Random DTX} in High Mobility Scenario]\label{problemoriginalhm}
{\setlength{\arraycolsep}{0.0em}
\begin{eqnarray*}
{q}_{\mathrm{hm}}^* \triangleq\mathop {\max }\limits_{\mathbf{T},\beta} &&\;\;q_{\mathrm{hm}}(\mathbf{T},\beta)\\
 s.t.\;\;
&&(\ref{eqconstcachingprob}),(\ref{eqconst2mbcachesize}),\beta\in(0,1],
\end{eqnarray*}\setlength{\arraycolsep}{5pt}}where ${q}_{\mathrm{hm}}^*={q}_{\mathrm{hm}}(\mathbf{T}^*,\beta^*)$ denotes an optimal value and $(\mathbf{T}^*,\beta^*)$ denotes {an} optimal solution.
\end{Problem}

{Problem \ref{problemoriginalhm} maximizes a non-concave function over a convex set, and hence is non-convex. In general, it is difficult to obtain a globally optimal solution of a non-convex problem. By exploring properties of the objective function $q_{\mathrm{hm}}(\mathbf{T},\beta)$, in the following, we can obtain a globally optimal solution of Problem \ref{problemoriginalhm}.}

Recall that Lemma \ref{propMonotonicityeqCorollaryphiequalximb} shows that $q_{n,\mathrm{hm}}(\mathbf{T}_n,\beta)$ increases with $\beta$ for all $\mathbf{T}_n$. Thus, we know that $\beta^*=1$. It remains to obtain $\mathbf{T}^*$ by maximizing $q_{n,\mathrm{hm}}(\mathbf{T}_n,1)$ w.r.t. $\mathbf{T}$, i.e., solving the following problem.
\begin{Problem}[{Optimization of Random Caching} in High Mobility Scenario]\label{problemoriginalhmeqv}
{\setlength{\arraycolsep}{0.0em}
\begin{eqnarray*}
{q}_{\mathrm{hm}}^* =\mathop {\max }\limits_{\mathbf{T}} &&\;\;{q}_{\mathrm{hm}}(\mathbf{T},1) \\
 s.t.\;\;
&&(\ref{eqconstcachingprob}),(\ref{eqconst2mbcachesize}).
\end{eqnarray*}\setlength{\arraycolsep}{5pt}}
\end{Problem}

Recall that Lemma \ref{propMonotonicityeqCorollaryphiequalximb} shows that $q_{n,\mathrm{hm}}(\mathbf{T}_n,\beta)$ is a concave function of $\mathbf{T}_n$, implying {that} ${q}_{\mathrm{hm}}(\mathbf{T},\beta)$ in (\ref{eqCorollaryphiequalxi2mb}) is a concave function of $\mathbf{T}$. Thus, Problem \ref{problemoriginalhmeqv} is a convex optimization problem and can be efficiently solved by the  interior point method.  {Consider} a special case of $C_k=C$ for all $k\in\mathcal{K}$, i.e., equal cache size across { all tiers}.  {U}sing KKT conditions, the optimal solution of Problem  \ref{problemoriginalhmeqv} can be characterized as follows.
\begin{Lemma}[Optimal Solution of Problem  \ref{problemoriginalhmeqv} {When $C_k=C$, $k\in\mathcal{K}$}]\label{lemmaoptimalsolutionHMingeneralcase}
When $C_k=C$ for all $k\in\mathcal{K}$, an optimal solution $\mathbf{T}^*$ of Problem  \ref{problemoriginalhmeqv} is given by\footnote{Note that, {for all $k\in\mathcal{K}$,} $T_{n,k}=g^{-1}(\eta^*)$ can be obtained by solving $g(T_{n,k}^*)=\eta$ using the bisection method, and $\eta^*$ can be obtained  by solving $\sum_{n\in\mathcal{N}}T_{n,k}^*=C$ using the bisection method. }
{\begingroup\makeatletter\def\f@size{12}\check@mathfonts\def\maketag@@@#1{\hbox{\m@th\normalsize\normalfont#1}}\setlength{\arraycolsep}{0.0em}
\begin{eqnarray*}\label{eqeqTfkHMgc}
T_{n,k}^*=\begin{cases}
                 0, & \mbox{if } \eta^*\ge  g(0), \\
                 1, & \mbox{if } \eta^*\le  g(1), \\
                 g^{-1}(\eta^*), & \mbox{otherwise},
            \end{cases}\quad n\in\mathcal{N},k\in\mathcal{K},
\end{eqnarray*}\setlength{\arraycolsep}{5pt}\endgroup}where $g^{-1}(\cdot)$ denotes the inverse function of function $g(\cdot)$, given by
{\setlength{\arraycolsep}{0.0em}
\begin{eqnarray*}\label{eqTfkequalityHMgc}
g(x)\triangleq a_n M\left(1-\frac{x}{xW(1)+(1-x)V(1)}\right)^{M-1}\frac{V(1)}{\left(xW(1)+(1-x)V(1)\right)^2},
\end{eqnarray*}\setlength{\arraycolsep}{5pt}}and $\eta^*$ satisfies $\sum_{n\in\mathcal{N}}T_{n,k}^*=C$.
\end{Lemma}

{Based on $\beta^*=1$ and an optimal solution of Problem \ref{problemoriginalhmeqv}, we can obtain a globally optimal solution of Problem \ref{problemoriginalhm}.}

\section{Static Scenario}
In this section, we consider the static  scenario. {We first analyze the STP and then} maximize the STP by optimizing the design parameters of random caching and random DTX.\vspace{-3mm}
\subsection{Performance Analysis}
In this part, we  {analyze the STP in the general SIR threshold regime and the low SIR threshold regime,} respectively. To be specific, we only need to analyze the STP of file $n$, i.e., $q_{n,\mathrm{st}}(\mathbf{T}_n,\beta)$, in the two regimes. Then, by (\ref{eqDefphihmaverage}), we can directly obtain the STP, i.e., ${q}_{\mathrm{st}}(\mathbf{T},\beta)$.
\subsubsection{Performance Analysis in General SIR Threshold Regime}
In this part, we analyze {$q_{n,\mathrm{st}}(\mathbf{T}_n,\beta)$} in the general SIR threshold regime, using tools from stochastic geometry. It is challenging to calculate $q_{n,\mathrm{st}}(\mathbf{T}_n,\beta)$, {as $\mathrm{SIR}_{n,0}(t)$, $t=1,2,\cdots,M$} are correlated. To address this challenge, by using the binomial expansion theorem, we first rewrite $q_{n,\mathrm{st}}(\mathbf{T}_n,\beta)$ in (\ref{eqDefphist}) as $q_{n,\mathrm{st}}(\mathbf{T}_n,\beta)=\sum_{m=1}^{M}\binom{M}{m} (-1)^{m+1} \mathbbm{E}_{\Phi}\left(\left( \Pr\left(\mathrm{SIR}_{n,0}(t)>\theta|\Phi\right)\right)^m\right)$. Note that, conditioned on $\Phi$, $\mathrm{SIR}_{n,0}(t)$, $t=1,2,\cdots,M$ are i.i.d.. Thus, we {can first} analyze the distribution of $\mathrm{SIR}_{n,0}(t)$, conditioned on $\Phi${, and then derive $q_{n,\mathrm{st}}(\mathbf{T}_n,\beta)$ by deconditioning on $\Phi$.} Thus, we have the following theorem.
\begin{Theorem}[STP in Static Scenario]\label{TheoremSDP}
The STP of file $n$ in the static scenario is given by,
{\setlength{\arraycolsep}{0.0em}
\begin{equation}\label{eqCorollaryphiequalxi}
q_{n,\mathrm{st}}(\mathbf{T}_n,\beta)=\sum_{m=1}^{M}\binom{M}{m}\frac{(-1)^{m+1}{\beta^m\sum_{k\in\mathcal{K}}z_kT_{n,k}}}{F_m(\beta)\sum_{k\in\mathcal{K}}z_kT_{n,k}  + G_m(\beta)(1-\sum_{k\in\mathcal{K}}z_kT_{n,k})},\quad n\in\mathcal{N},
\end{equation}
\setlength{\arraycolsep}{5pt}}where $z_k$ is given by Theorem \ref{TheoremSDPMb}, $F_m(\beta)$ and $G_m(\beta)$ are given, respectively, by
{\setlength{\arraycolsep}{0.0em}
\begin{eqnarray}
  F_m(\beta)&=&\sum_{i=0}^{m}\binom{m}{i}\beta^i\left(1-\beta\right)^{m-i}{_2F_1}\left(-\frac{2}{\alpha},i;1-\frac{2}{\alpha};-\theta\right),\label{eqFm}\\
  G_m(\beta)&=&\sum_{i=0}^{m}\binom{m}{i}\beta^i\left(1-\beta\right)^{m-i}\frac{\Gamma\left(i+\frac{2}{\alpha}\right)}{\Gamma(i)}\Gamma\left(1-\frac{2}{\alpha}\right)\theta^{\frac{2}{\alpha}}.\label{eqGm}
\end{eqnarray}\setlength{\arraycolsep}{5pt}}
\end{Theorem}

\indent\indent \textit{Proof}: See Appendix \ref{ProofTheoremSDP}. $\hfill\blacksquare$

\begin{figure*}[!t]%
    \centering
        \subfloat[$T_{n,1}=0.8$ and $\beta=0.9$.]{\includegraphics[width=3.3in]{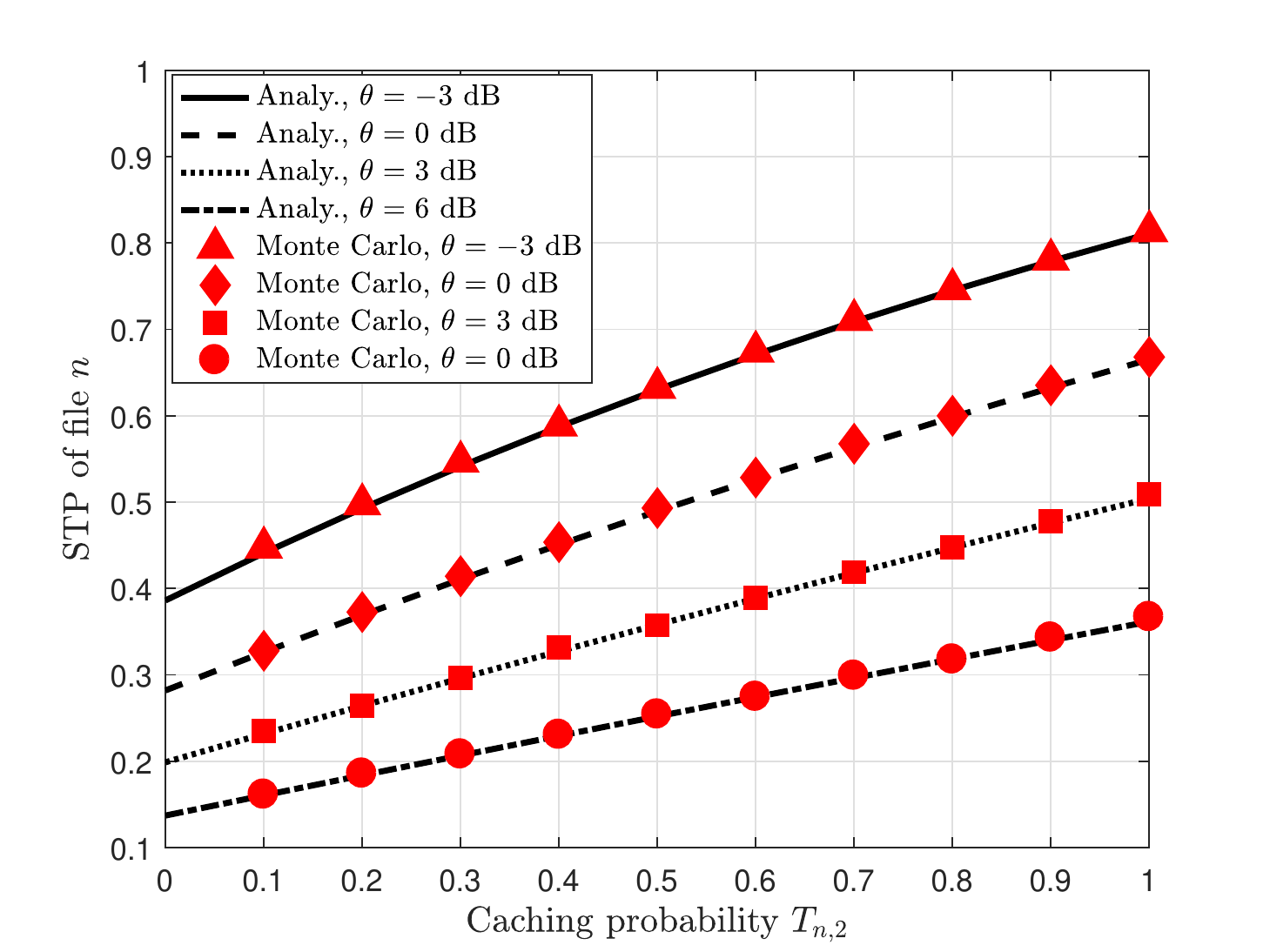}}
        \subfloat[$T_{n,1}=T_{n,2}=0.8$.]{\includegraphics[width=3.3in]{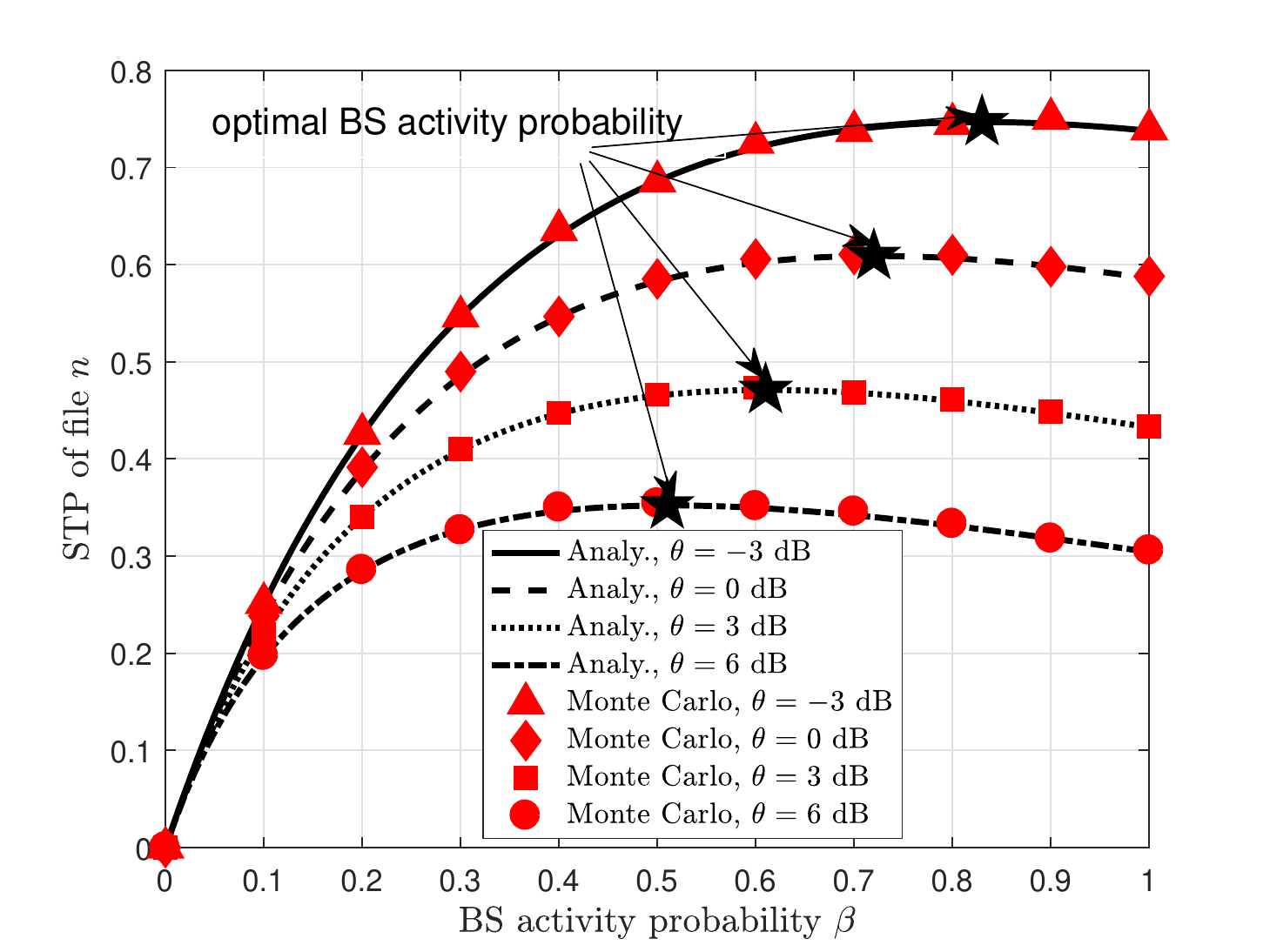}}

      \caption{{$q_{n,\mathrm{st}}(\mathbf{T}_n,\beta)$} versus $T_{n,k}$ and $\beta$, respectively, in the static scenario. The simulation parameters are given in Fig. \ref{figsdpasymptoticZero}.}\label{figsdpingeneralregionST}\vspace{-0.8cm}
\end{figure*}

Similarly to Theorem \ref{TheoremSDPMb}, Theorem \ref{TheoremSDP} provides a closed-form expression for {$q_{n,\mathrm{st}}(\mathbf{T}_n,\beta)$} in the general SIR threshold regime. However, the expression in Theorem \ref{TheoremSDP} is more complex than that in Theorem \ref{TheoremSDPMb}{, due to the correlations across the $M$ consecutive slots.} Fig. \ref{figsdpingeneralregionST} plots {$q_{n,\mathrm{st}}(\mathbf{T}_n,\beta)$}  versus  {$T_{n,k}$} and {$\beta$}, respectively, in the static scenario, verifying Theorem \ref{TheoremSDP}. It is worth noting that unlike the high mobility scenario, it is hard to analytically characterize {how $q_{n,\mathrm{st}}(\mathbf{T}_n,\beta)$ changes with $T_{n,k}$ and $\beta$}. However, from Fig. \ref{figsdpingeneralregionST}, we can observe some properties in the static scenario for the considered setup. Specifically, Fig. \ref{figsdpingeneralregionST}(a) shows that {$q_{n,\mathrm{st}}(\mathbf{T}_n,\beta)$ is an increasing function of $T_{n,k}$, for all $k\in\mathcal{K}$}, which reveals the advantage of caching. {Fig. \ref{figsdpingeneralregionST}(b) shows that in the static scenario, for some $\mathbf{T}_n$, there exists an optimal BS activity probability $\beta^*\le 1$ that maximizes $q_{n,\mathrm{st}}(\mathbf{T}_n,\beta)$, in contrast to the case in the high mobility scenario, {where   $\beta^*=1$} for any given $\mathbf{T}_n$. {Although the penalty of random DTX in signal reduction may overtake its advantage in interference reduction in one particular slot (see (\ref{eqSTPOneSlot})), random DTX is also able to reduce interference correlation across different slots in the static scenario \cite{ManagingInterferenceCorrelationZhongYi}. When $\beta>\beta^*$, the advantages of random DTX outweigh its penalty and when $\beta<\beta^*$, its penalty outweighs its advantages.}
}

\subsubsection{Performance Analysis in Low SIR Threshold Regime}

\begin{figure*}[!t]
    \centering
        {\includegraphics[width=3.3in]{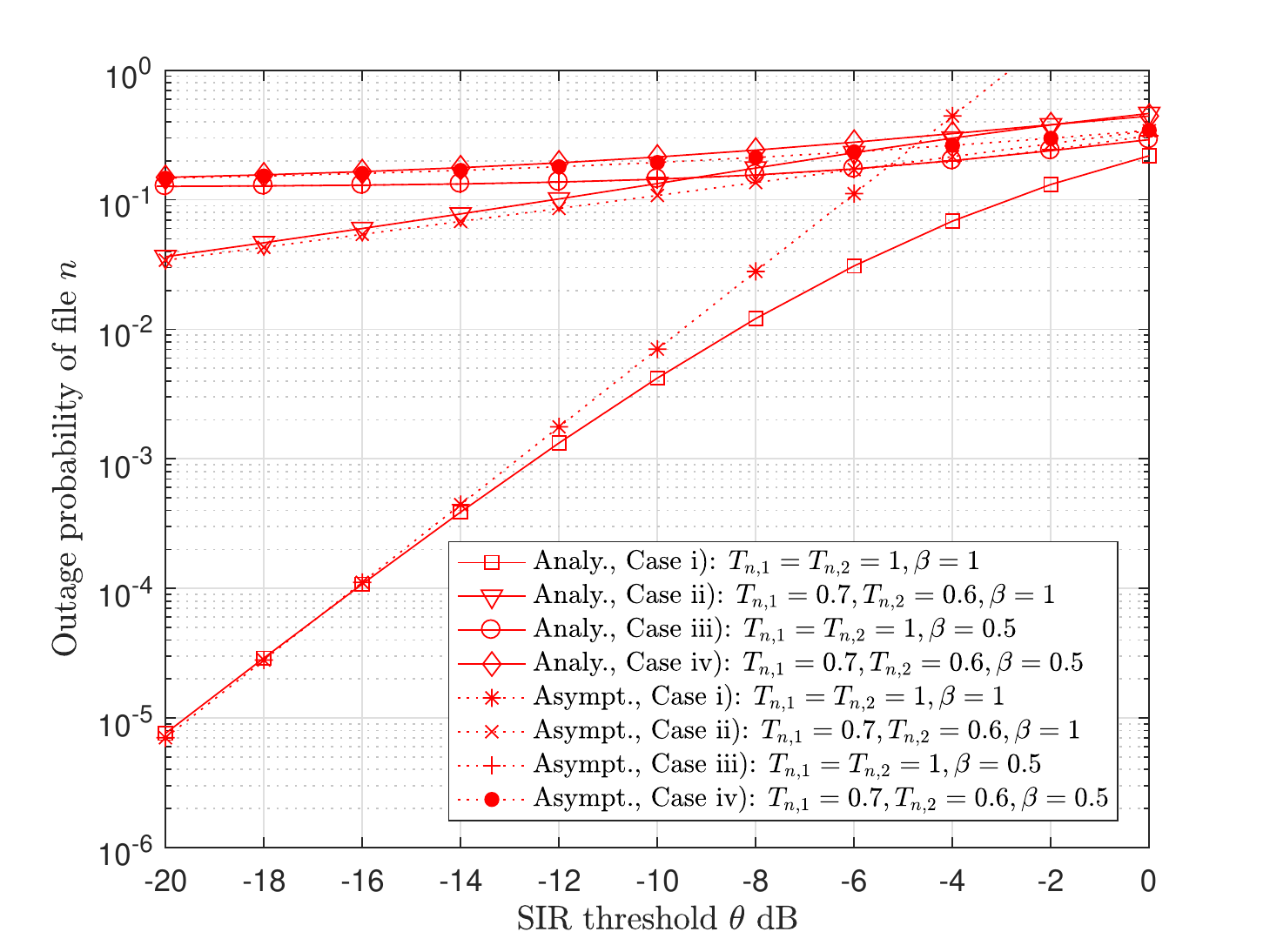}}

      \caption{{$\bar{q}_{n,\mathrm{st}}(\mathbf{T}_n,\beta)$} versus $\theta$ in the static scenario in different cases. $P_1=20$ W, $P_2=0.13$ W, $\lambda_1=\frac{1}{ 250^2\pi}$, $\lambda_2=\frac{1}{ 50^2\pi}$, $\alpha=4$, and $M=3$.}\label{figsdpexactandasymptosST}\vspace{-0.8cm}
\end{figure*}

To further obtain insights, in this part, we {analyze the outage probability of file $n$ which is defined as $\bar{q}_{n,\mathrm{st}}(\mathbf{T}_n,\beta)\triangleq 1- q_{n,\mathrm{st}}(\mathbf{T}_n,\beta)$,} in  the low SIR threshold regime, i.e., $\theta\rightarrow0$. By considering the same four cases as in Lemma~\ref{lemmathetaarrow0mb}, we have the following result.
\begin{Lemma}[Outage Probability in Static  Scenario When $\theta\to0$]\label{lemmathetaarrow0}
In the static scenario, when $\theta\rightarrow0$, we have
{\setlength{\arraycolsep}{0.0em}
\begin{eqnarray}\label{eqlemmaphifasyTf}
\bar{q}_{n,\mathrm{st}}(\mathbf{T}_n,\beta) \sim \left(1-\beta\right)^M+c_{\mathrm{st}}(\mathbf{T}_n,\beta) ,\quad n\in\mathcal{N},
\end{eqnarray}\setlength{\arraycolsep}{5pt}}where
{\setlength{\arraycolsep}{0.0em}
\begin{eqnarray}\label{eqcst}
c_{\mathrm{st}}(\mathbf{T}_n,\beta)\triangleq \begin{cases}
\theta^M\frac{\partial^M}{\partial x^M}\left({_1F_1}(-\frac{2}{\alpha};1-\frac{2}{\alpha};x)\right)^{-1}|_{x=0}, & \mbox{case i)},\\
\theta^{\frac{2}{\alpha}}\left(\frac{1}{\sum_{k\in\mathcal{K}}z_kT_{n,k}}-1\right)\Gamma\left(1-\frac{2}{\alpha}\right) \sum_{m=1}^{M}\binom{M}{m}\\ \;\times(-1)^{m+1}\beta^m\sum_{i=0}^{m}\binom{m}{i} \beta^i(1-\beta)^{m-i}\frac{\Gamma\left(i+\frac{2}{\alpha}\right)}{\Gamma(i)}, &\mbox{cases ii and iv)},\\
        \theta(1-\beta)^{M-1}M\beta^2\frac{2}{\alpha-2}, &\mbox{case iii)}.
                           \end{cases}
\end{eqnarray}\setlength{\arraycolsep}{5pt}}Here, ${_1F_1}(a;b;x)$ is the confluent hypergeometric function of the first kind.
\end{Lemma}

\indent\indent \textit{Proof}: See Appendix \ref{Prooflemmathetaarrow0}. $\hfill\blacksquare$

From Lemma \ref{lemmathetaarrow0}, we can see that both the caching probability $\mathbf{T}_n$ and the BS activity probability $\beta$ significantly affect the asymptotic behaviours of $\bar{q}_{n,\mathrm{st}}(\mathbf{T}_n,\beta)$ when $\theta\to 0$, but in  different manners. Fig. \ref{figsdpexactandasymptosST} plots {$\bar{q}_{n,\mathrm{st}}(\mathbf{T}_n,\beta)$ versus $\theta$} and indicates that Lemma \ref{lemmathetaarrow0} provides a good approximation for $\bar{q}_{n,\mathrm{st}}(\mathbf{T}_n,\beta)$ when $\theta$ is small.

Similarly{, we further characterise the diversity gain in the static scenario, i.e.,}
{\setlength{\arraycolsep}{0.0em}
\begin{eqnarray}\label{eqdefdiversitygain}
d_{\mathrm{st}}=\lim_{\theta\rightarrow0}\frac{\log\left(\bar{q}_{n,\mathrm{st}}(\mathbf{T}_n,\beta)\right)}{\log\theta}.
\end{eqnarray}\setlength{\arraycolsep}{5pt}}{From Lemma \ref{lemmathetaarrow0}, we have the following result.}
\begin{Lemma}[Diversity Gain in Static Scenario]\label{lemmaDiversitygain}
The diversity gain in the static scenario is given~by
{\setlength{\arraycolsep}{0.0em}
\begin{eqnarray}\label{eqdiversityorderSTL1}
d_{\mathrm{st}}=\begin{cases}
                  M, & \mbox{case i)},\\
                  \frac{2}{\alpha}, & \mbox{case ii)},\\
                  0, &\mbox{cases iii) and iv)}.
                \end{cases}
\end{eqnarray}\setlength{\arraycolsep}{5pt}}
\end{Lemma}

{ Lemma \ref{lemmaDiversitygain} can be interpreted in the same way as Lemma \ref{lemmaDiversitygainmb}. Comparing the diversity gains for case ii) in Lemma \ref{lemmaDiversitygain} and Lemma \ref{lemmaDiversitygainmb}, we know that for case ii), the diversity gain  in the high mobility scenario is $M$ times of that in the static scenario due to  the fact that  there is  no interference correlation in the high mobility scenario.

}

\subsection{Performance Optimization}
By substituting (\ref{eqCorollaryphiequalxi}) into (\ref{eqDefphist}), the STP in the static scenario is calculated as
{\setlength{\arraycolsep}{0.0em}
\begin{equation}\label{eqCorollaryphiequalxi2}
{q}_{\mathrm{st}}(\mathbf{T},\beta)=\sum_{n\in\mathcal{N}}a_n \sum_{m=1}^{M}\binom{M}{m}\frac{(-1)^{m+1}{\beta^m\sum_{k\in\mathcal{K}}z_kT_{n,k}}}{F_m(\beta)\sum_{k\in\mathcal{K}}z_kT_{n,k}  + G_m(\beta)(1-\sum_{k\in\mathcal{K}}z_kT_{n,k})}.
\end{equation}
\setlength{\arraycolsep}{5pt}}The caching distribution $\mathbf{T}$ and BS activity probability $\beta$ significantly affects the STP in the static scenario. We would like to maximize ${q}_{\mathrm{st}}(\mathbf{T},\beta)$ in (\ref{eqCorollaryphiequalxi2}) by jointly optimizing $\mathbf{T}$ and $\beta$. Specifically, we have the following optimization problem.
\begin{Problem}[{Optimization of} Random Caching and Random DTX  in Static Scenario]\label{problemoriginalst}
{\setlength{\arraycolsep}{0.0em}
\begin{eqnarray}
{q}_{\mathrm{st}}^* \triangleq\mathop {\max }\limits_{\mathbf{T},\beta} &&\;\;{q}_{\mathrm{st}}(\mathbf{T},\beta)\nonumber\\
 s.t.\;\;
&&(\ref{eqconstcachingprob}),(\ref{eqconst2mbcachesize}),\beta\in(0,1],\label{eqAllConstr}
\end{eqnarray}\setlength{\arraycolsep}{5pt}}where ${q}_{\mathrm{st}}^*={q}_{\mathrm{st}}(\mathbf{T}^*,\beta^*)$ denotes the optimal value and $(\mathbf{T}^*,\beta^*)$ denotes the optimal solution.
\end{Problem}

Problem \ref{problemoriginalst} maximizes a non-concave function over a convex set, and hence is non-convex in general. { Recall that in Section \ref{subsectionPeromanceMetric}, when $M=1$, we have $q_{n,\mathrm{hm}}=q_{n,\mathrm{st}}$, implying that {when $M=1$,} Problem~\ref{problemoriginalst} can be solved by using the same method as {for} Problem \ref{problemoriginalhm}. Thus, in the following, we focus on solving Problem \ref{problemoriginalst} when $M\ge 2$.  Note that,} as ${q}_{\mathrm{st}}(\mathbf{T},\beta)$ is differentiable, in general, we can obtain a stationary point of Problem~\ref{problemoriginalst} {when $M\ge 2$,} using the gradient projection method with a diminishing stepsize.\footnote{Note that a stationary point is a point that satisfies the necessary optimality conditions of a non-convex optimization problem, and it is the classic goal in the design of iterative algorithms for non-convex optimization problems.} However, the rate of convergence of the gradient projection method is strongly dependent on the choices of stepsize. If {it is} chosen improperly, it may take a large number of iterations to meet some convergence criterion, especially when the number of variables in Problem \ref{problemoriginalst} is large. To address this issue,  we propose a more efficient algorithm to obtain a stationary point of Problem \ref{problemoriginalst}{, based on alternating optimization.} {Specifically, we partition the variables in Problem \ref{problemoriginalst} into two blocks, i.e., $ \mathbf{T}$ and $\beta$, and separate the constraint sets of these two blocks. Then, we solve a random caching optimization problem and a random DTX optimization problem alternatively.}

\subsubsection{Random Caching Optimization}\label{subsubrandomcaching}
First, we consider the optimization of {the random caching probability} $\mathbf{T}$ while fixing $\beta$.
\begin{Problem}[{Optimization of} Random Caching  for Given $\beta$]\label{subProb1}
{\setlength{\arraycolsep}{0.0em}
\begin{eqnarray*}
 \max_{\mathbf{T}} &&\;\;q_{\mathrm{st}}(\mathbf{T},\beta)\\
 s.t.\;\;
&&(\ref{eqconstcachingprob}),(\ref{eqconst2mbcachesize}).
\end{eqnarray*}\setlength{\arraycolsep}{5pt}}
\end{Problem}

To solve Problem \ref{subProb1}, we first analyze its {structural properties}. Let $\mathcal{M}_1$  and $\mathcal{M}_2$ denote the sets of all the odd and even numbers in the set $\{1,2,\cdots,M\}$, respectively. We rewrite {$q_{\mathrm{st}}(\mathbf{T},\beta)$ in} (\ref{eqCorollaryphiequalxi2}) as
{\setlength{\arraycolsep}{0.0em}
\begin{equation*}\label{eqCorollaryphiequalxi22}
q_{\mathrm{st}}(\mathbf{T},\beta) =q_1(\mathbf{T},\beta)-q_2(\mathbf{T},\beta),
\end{equation*}\setlength{\arraycolsep}{5pt}}where $q_i(\mathbf{T},\beta)$ is given by
{\setlength{\arraycolsep}{0.0em}
\begin{equation*}
q_i(\mathbf{T},\beta) = \sum_{n\in\mathcal{N}}a_n \sum_{m\in\mathcal{M}_i}\binom{M}{m} \frac{\beta^m{\sum_{k\in\mathcal{K}}z_kT_{n,k}}} {F_m(\beta)\sum_{k\in\mathcal{K}}z_kT_{n,k}  + G_m(\beta)(1-\sum_{k\in\mathcal{K}}z_kT_{n,k})}, \quad{i=1,2}.
\end{equation*}\setlength{\arraycolsep}{5pt}}It can be easily verified that $q_i(\mathbf{T},\beta)$ is a concave function of $\mathbf{T}$. {Thus,} Problem \ref{subProb1} is a difference-of-convex (DC) programming problem and can be solved based on the convex-concave procedure (CCP) {\cite{Lipp2016}}. The basic idea of the CCP is to linearize the convex terms of the objective function ({i.e., $-q_2(\mathbf{T},\beta)$ in $q_{\mathrm{st}}(\mathbf{T},\beta)$}) to obtain a concave objective for a maximization problem, {and then solve a sequence of convex problems successively.} Specifically, at iteration $j$, we solve the following problem:
{\setlength{\arraycolsep}{0.0em}
\begin{eqnarray}\label{eqsubproblem1DC}
\mathbf{T}^{(j)} &\triangleq  \arg \max\limits_{\mathbf{T}} \;\;q_1(\mathbf{T},\beta)-\tilde{q}_2(\mathbf{T},\beta;\mathbf{T}^{(j-1)}) \\
&s.t.\;\;(\ref{eqconstcachingprob}),(\ref{eqconst2mbcachesize}),\nonumber
\end{eqnarray}\setlength{\arraycolsep}{5pt}}where $\tilde{q}_2(\mathbf{T},\beta;\mathbf{T}^{(j-1)})\triangleq q_2(\mathbf{T}^{(j-1)},\beta)+ (\mathbf{T}-\mathbf{T}^{(j-1)})^T\triangledown_{\mathbf{T}}{q}_2(\mathbf{T}^{(j-1)},\beta)$, and
 $\triangledown_{\mathbf{T}}{q}_2(\mathbf{T}^{(j-1)},\beta)$ denotes the gradient of $q_2(\mathbf{T},\beta)$ at $\mathbf{T}=\mathbf{T}^{(j-1)}$.

Since $q_1(\mathbf{T},\beta)$ and $\tilde{q}_2(\mathbf{T},\beta;\mathbf{T}^{(j-1)})$ are concave and linear w.r.t. $\mathbf{T}$, respectively, {the optimization in (\ref{eqsubproblem1DC}) is a convex problem} which can be efficiently solved by the interior point method. The details of the proposed iterative algorithm are summarized in Algorithm \ref{algDC}. {Note that, it has} been shown in \cite{Lipp2016} that the sequence {$\{\mathbf{T}^{(j)}\}_{j=1}^{\infty}$} generated by Algorithm \ref{algDC} converges to a stationary point of Problem \ref{subProb1}.

\begin{algorithm}[!t]
\caption{Stationary Point of Problem \ref{subProb1} Based on CCP}
\begin{algorithmic}[1]\small
\STATE \textbf{Initialization:} set $j=0$ and choose any $\mathbf{T}^{(0)}$ satisfying (\ref{eqconstcachingprob}) and (\ref{eqconst2mbcachesize}).
\STATE \textbf{repeat}
\STATE Obtain stationary point $\mathbf{T}^{(j+1)}$ by solving the optimization in (\ref{eqsubproblem1DC}).
\STATE $j\leftarrow j+1$.
\STATE \textbf{until} convergence criterion is met.
\end{algorithmic}\label{algDC}
\end{algorithm}

Next, we consider a special case of $C_k=C$ for all $k\in\mathcal{K}$, i.e., equal cache size across all tiers. {In this case, the optimization in (\ref{eqsubproblem1DC}) is convex. Similarly,} using KKT conditions, {we can obtain an} optimal solution of the problem  in (\ref{eqsubproblem1DC})  as follows.
\begin{Lemma}[Optimal Solution of Problem  in (\ref{eqsubproblem1DC}) When $C_k=C$ for all $k\in\mathcal{K}$]\label{lemmaoptimalsolutionST}
When $C_k=C$ for all $k\in\mathcal{K}$, an optimal solution $\mathbf{T}^*$ of problem  in (\ref{eqsubproblem1DC}) is given by
{\begingroup\makeatletter\def\f@size{12}\check@mathfonts\def\maketag@@@#1{\hbox{\m@th\normalsize\normalfont#1}}\setlength{\arraycolsep}{0.0em}
\begin{eqnarray*}\label{eqeqTfkHMgc}
T_{n,k}^{(j)}=\begin{cases}
                 0, & \mbox{if } \eta^*\ge  f(0), \\
                 1, & \mbox{if } \eta^*\le f(1), \\
                 f^{-1}(\eta^*), & \mbox{otherwise},
            \end{cases}\quad n\in\mathcal{N},k\in\mathcal{K},
\end{eqnarray*}\setlength{\arraycolsep}{5pt}\endgroup}where $f^{-1}(\cdot)$ denotes the inverse function of function $f(\cdot)$, given by
{\begingroup\makeatletter\def\f@size{11}\check@mathfonts
\def\maketag@@@#1{\hbox{\m@th\normalsize\normalfont#1}}\setlength{\arraycolsep}{0.0em}
\begin{eqnarray*}\label{eqTfkequalityHMgc}
f(x)&\triangleq&  a_n \sum_{m\in\mathcal{M}_1}\binom{M}{m}\frac{\beta^mG_m(\beta)}{\left(F_m(\beta) x  + G_m(\beta)(1- x)\right)^2}- a_n\sum_{m\in\mathcal{M}_2}\binom{M}{m}\frac{\beta^mG_m(\beta)}{\left(F_m(\beta) T_{n,k}^{(j-1)}  + G_m(\beta)(1- T_{n,k}^{(j-1)})\right)^2},
\end{eqnarray*}\setlength{\arraycolsep}{5pt}\endgroup}and $\eta^*$ satisfies $\sum_{n\in\mathcal{N}}T_{n,k}^*=C$.
\end{Lemma}


\subsubsection{Random DTX Optimization}\label{subsubsectionbetaST}
Next, we consider the optimization {of the BS activity probability} $\beta$ while fixing $\mathbf{T}$.
\begin{Problem}[Optimization of Random DTX  for Given $\mathbf{T}$]\label{subProb2}
{\setlength{\arraycolsep}{0.0em}
\begin{eqnarray*}
 \max_{\beta} &&\;\;q_{\mathrm{st}}(\mathbf{T},\beta)\\
 s.t.\;\;
&&\beta\in(0,1].
\end{eqnarray*}\setlength{\arraycolsep}{5pt}}
\end{Problem}

When $M=2$, it can be easily verified that Problem \ref{subProb2} is convex and thus can be efficiently solved by the interior point method.  When $M\ge 3$, it is hard to determine the convexity of  Problem \ref{subProb2}. Since $q_{\mathrm{st}}(\mathbf{T},\beta)$ is a continuously differentiable function of $\beta$, {a stationary point of Problem \ref{subProb2} can be efficiently obtained} by the gradient projection method.

\subsubsection{Alternating Optimization Procedure}

Based on the results in {Section \ref{subsubrandomcaching} and Section \ref{subsubsectionbetaST},} we develop an alternating optimization procedure for Problem \ref{problemoriginalst}, as summarized in Algorithm \ref{algSS}. {If the sequence $\{(\mathbf{T}^{(i)},\beta^{(i)})\}_{i=1}^{\infty}$ generated by Algorithm~\ref{algSS} is convergent, then every limit point of $\{(\mathbf{T}^{(i)},\beta^{(i)})\}_{i=1}^{\infty}$ is a stationary point of Problem~\ref{problemoriginalst}.}

\begin{algorithm}[!t]
\caption{Alternating Optimization Algorithm of Problem \ref{problemoriginalst}}
\begin{algorithmic}[1]\small
\STATE \textbf{Initialization:} set $i=0$ and choose any $\beta^{(0)}\in(0,1]$.
\STATE \textbf{repeat}
\STATE Fix $\beta^{(i)}$, and obtain a stationary point $\mathbf{T}^{(i)}$ of Problem \ref{subProb1} using Algorithm \ref{algDC}.
\STATE Fix $\mathbf{T}^{(i)}$, and obtain an optimal solution $\beta^{(i+1)}$ of  Problem~\ref{subProb2} when $M=2$ using  the interior point method or a stationary point $\beta^{(i+1)}$ of  Problem~\ref{subProb2}  when $M\ge 3$ using the gradient projection method.
\STATE $i\leftarrow i+1$.
\STATE \textbf{until} convergence criterion is met.
\end{algorithmic}\label{algSS}
\end{algorithm}



\begin{figure*}[!t]
    \centering
        \subfloat[STP versus iterations.]{\includegraphics[width=3.3in]{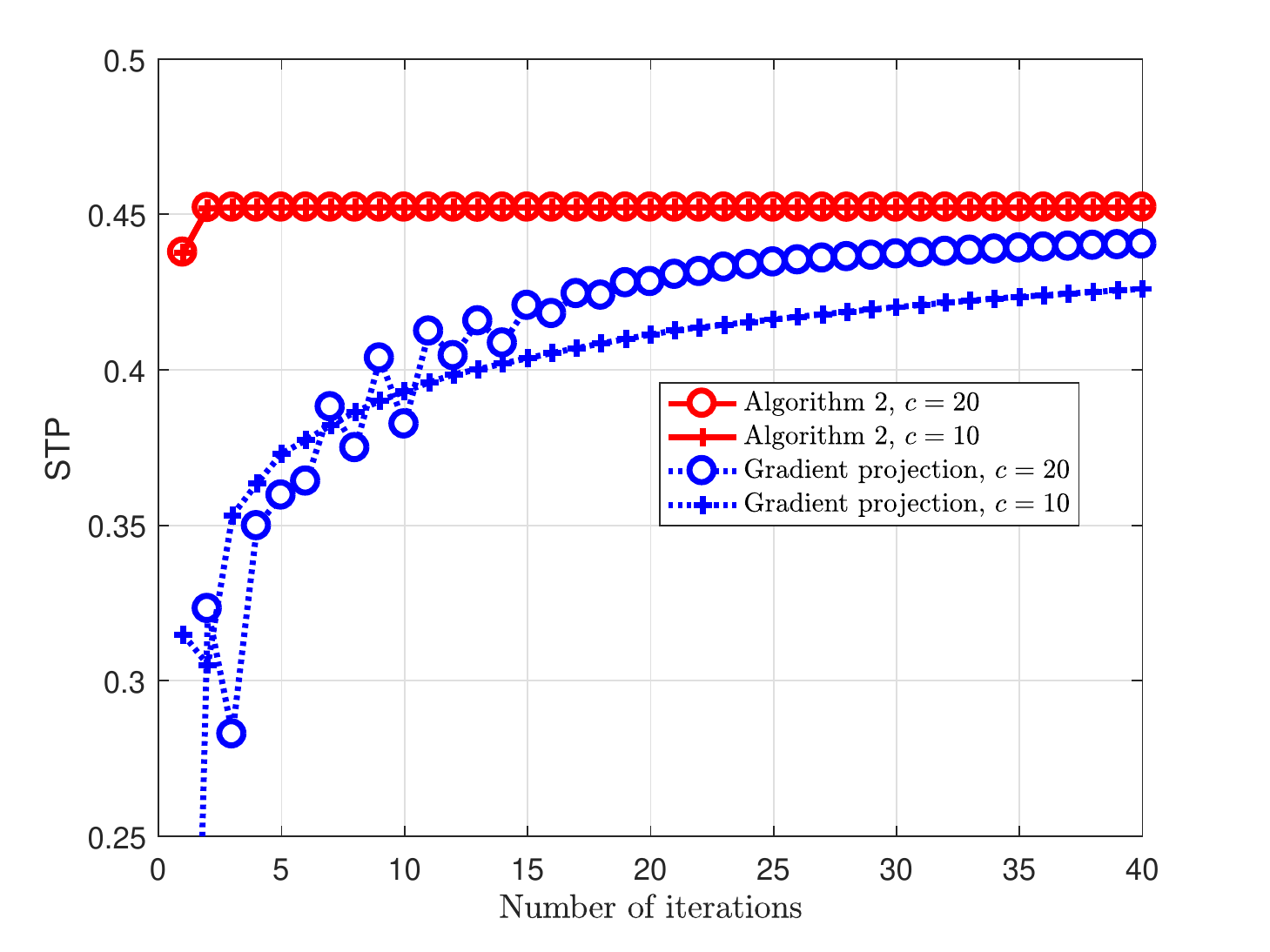}}
        \subfloat[Computing time versus  $C_2$ ($C_1$), $C_1=C_2+5$.]{\includegraphics[width=3.3in]{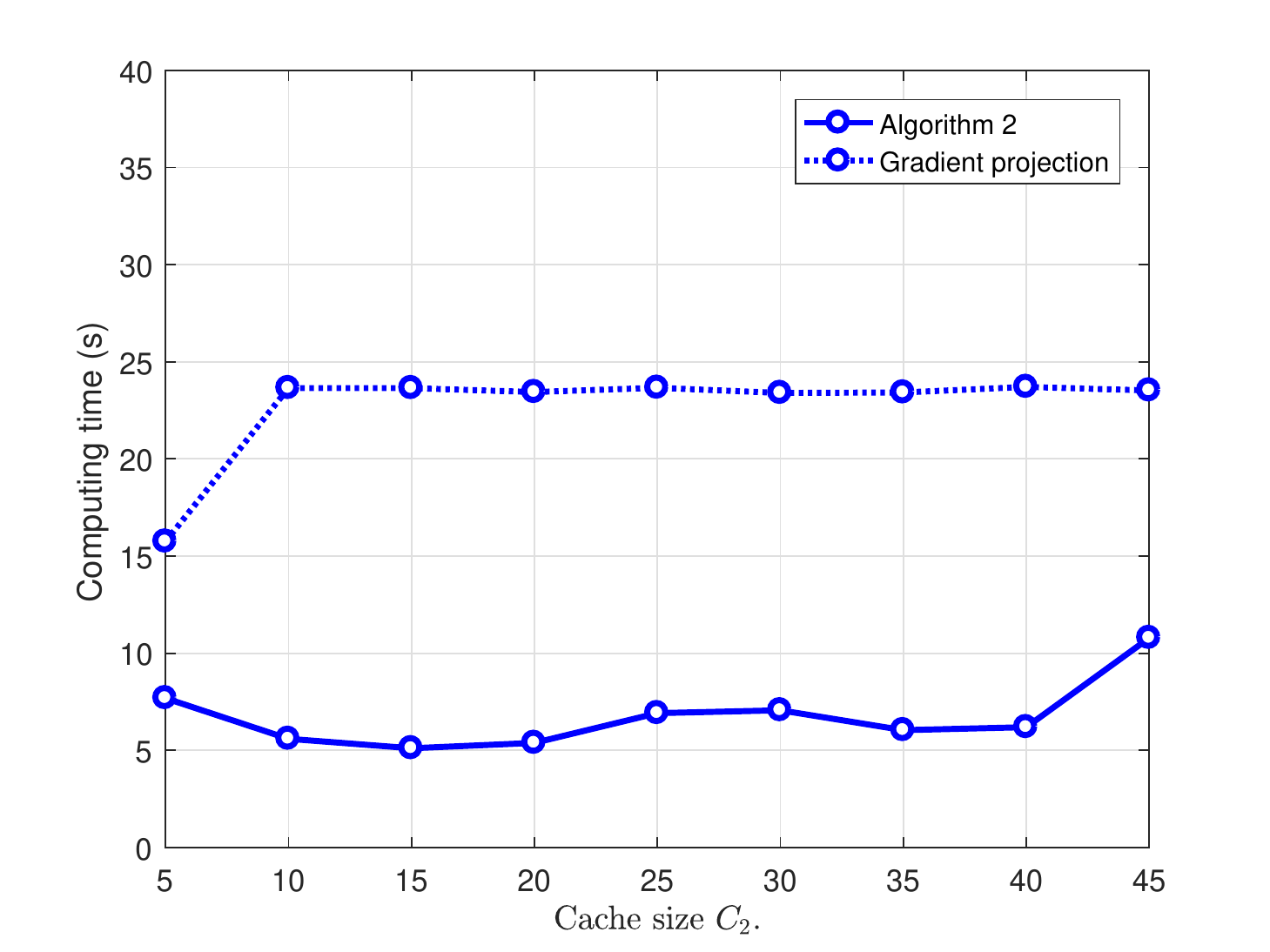}}

      \caption{Convergence rate and computing complexity of Algorithm \ref{algSS} at $M=3$. For the gradient projection method, we choose the stepsize {at iteration $i$} as $\epsilon(i)=\frac{c}{2+i^{0.55}}$. Note that, in (b), each point corresponds to the minimum computing time by choosing the optimal parameter $c \in \{5, 10, 15, 20, 25\}$. $P_1=20$ W, $P_2=0.13$ W, $\lambda_1=\frac{1}{ 250^2\pi}$, $\lambda_2=\frac{1}{ 50^2\pi}$, $\alpha=4$, $\theta=3$ dB, $C_1=25$, $C_2=15$, $N=50$, and $a_n=\frac{n^{-\gamma}}{\sum_{n \in\mathcal{N}}n^{-\gamma }}$, where $\gamma=0.8$ is the Zipf exponent.}\label{figConvergenceComputingTime}
\end{figure*}

Fig. \ref{figConvergenceComputingTime} {compares Algorithm \ref{algSS} and the gradient projection method for solving Problem \ref{problemoriginalst} in terms of the  convergence rate and computational  complexity.} From Fig. \ref{figConvergenceComputingTime}(a), we can see that {Algorithm \ref{algSS} is convergent} and the rate of convergence of the gradient projection method is strongly dependent on the choices of stepsize. In {contrast}, Algorithm \ref{algSS} has robust convergence performance. {In addition,} from Fig. \ref{figConvergenceComputingTime}(b), we can see that the  computing time of Algorithm  \ref{algSS} is shorter than the gradient projection method. {These demonstrate the advantage of Algorithm \ref{algSS} over the gradient projection method.}

\section{Numerical Results}
In this section, we first illustrate {the proposed  solutions in the high mobility and static scenarios. Then}, we compare the performance of the proposed solutions with some baselines in both scenarios. In the simulations, we consider a two-tier HetNet, i.e., $K=2$, consisting of a macrocell network as the 1st tier overlaid with a picocell network as the 2nd tier. Unless otherwise stated, the simulation settings are as follows: $P_1=20$ W, $P_2=0.13$ W, $\lambda_1=\frac{1}{ 250^2\pi}$, $\lambda_2=\frac{1}{ 50^2\pi}$, $\alpha=4$, $\theta=3$ dB, $C_1=25$, $C_2=15$, $N=50$, and $a_n=\frac{n^{-\gamma}}{\sum_{n \in\mathcal{N}}n^{-\gamma }}$, where $\gamma=0.8$ is the Zipf exponent.

\subsection{Proposed Solutions}

\begin{figure*}[!t]
    \centering
        \subfloat[Proposed caching probability  versus $n$.]{\includegraphics[width=3.3in]{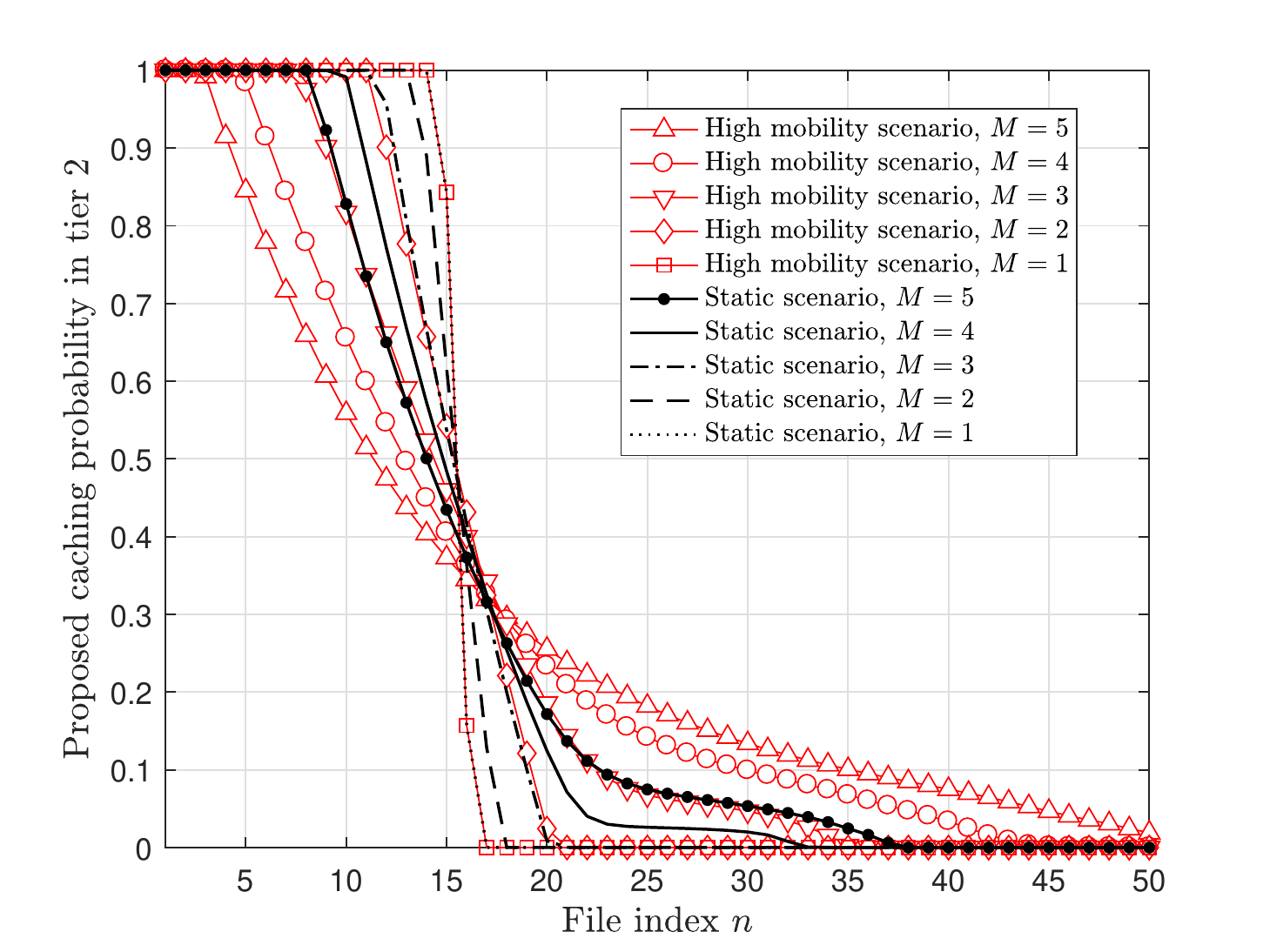}}
        \subfloat[Proposed BS activity probability versus $M$.]{\includegraphics[width=3.3in]{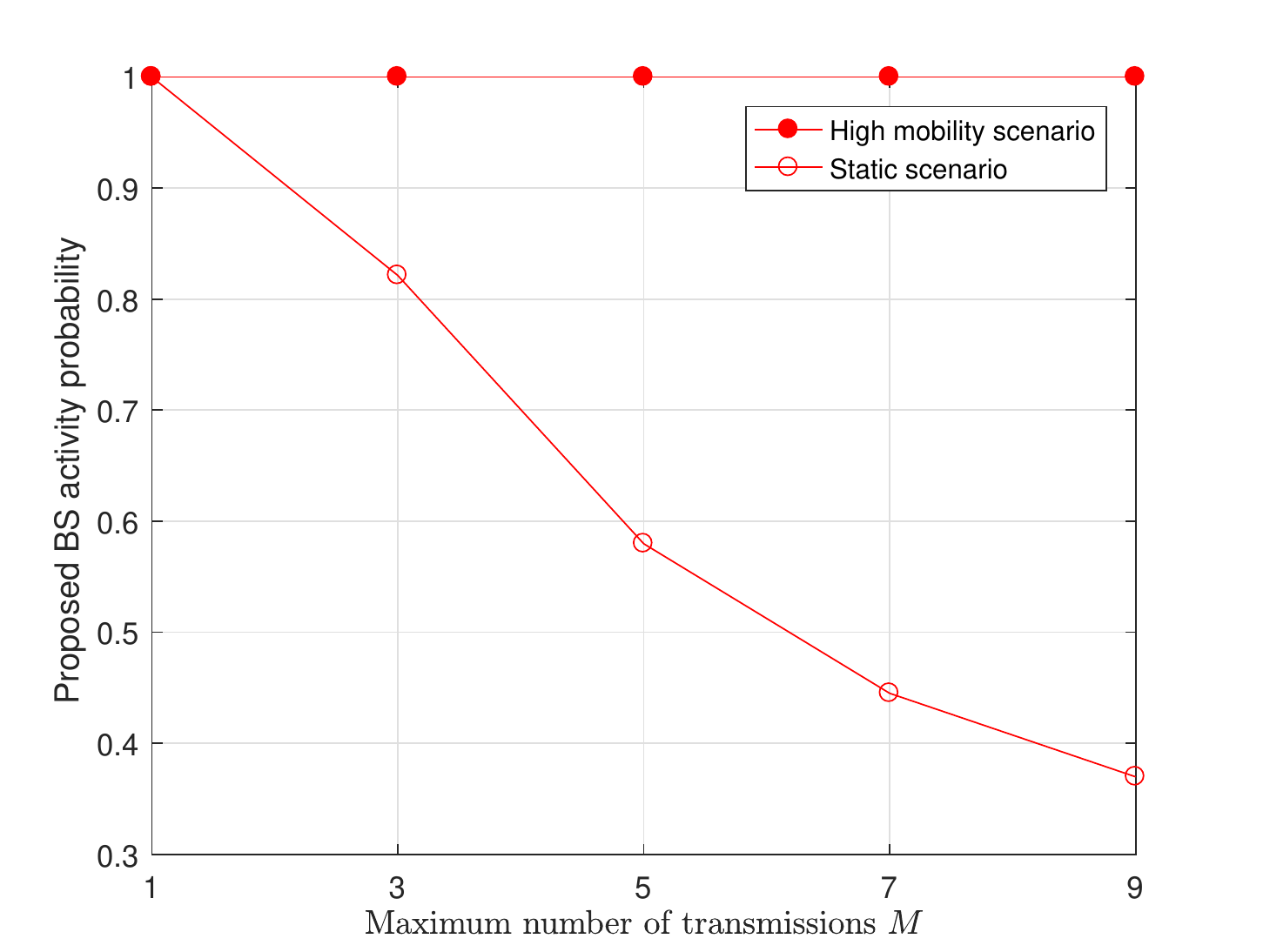}}

      \caption{Proposed caching probability in tier 2 and BS activity probability in the high mobility and static scenarios at $\theta=3$~dB.}\label{optSDPvsthetainHMST}
\end{figure*}

Fig. \ref{optSDPvsthetainHMST}(a) plots the {proposed} caching probability in tier 2 versus file index $n$ in the {two} scenarios. We can see that a {more popular} file corresponds to a larger caching probability, which is consistent with intuition. In addition,   {as} the maximum number of transmissions $M$ {increases},  more files can be stored at BSs, implying that retransmissions have a positive effect on the spatial file diversity. Fig. \ref{optSDPvsthetainHMST}(b) plots the {proposed} BS activity probability versus $M$ in the two scenarios. {We} can see that in the high mobility scenario, the optimal BS activity probability is $\beta^*=1${, verifying}  Lemma \ref{propMonotonicityeqCorollaryphiequalximb}.
{In the static scenario, we can see that $\beta^*$ decreases with $M$, which means that the larger the number of transmissions the more BSs should be silenced in one slot. This is because with $M$ increasing, {smaller interference correlation} can compensate lower BS availability.  }

\subsection{Comparisons between Proposed {Solutions} and Baselines}

\begin{figure*}[!t]
    \centering
        \subfloat[High mobility scenario.]{\includegraphics[width=3.3in]{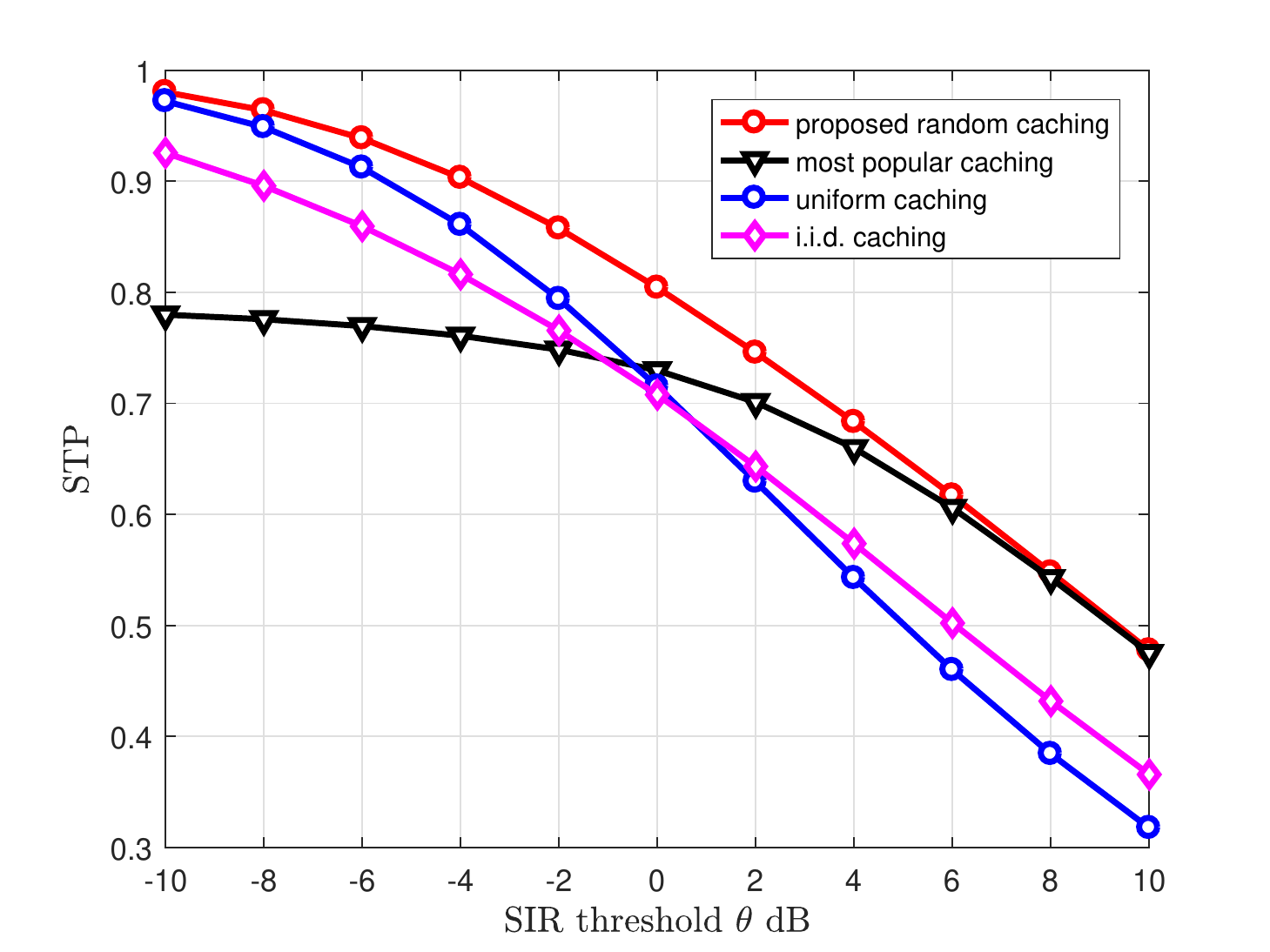}}
        \subfloat[Static scenario.]{\includegraphics[width=3.3in]{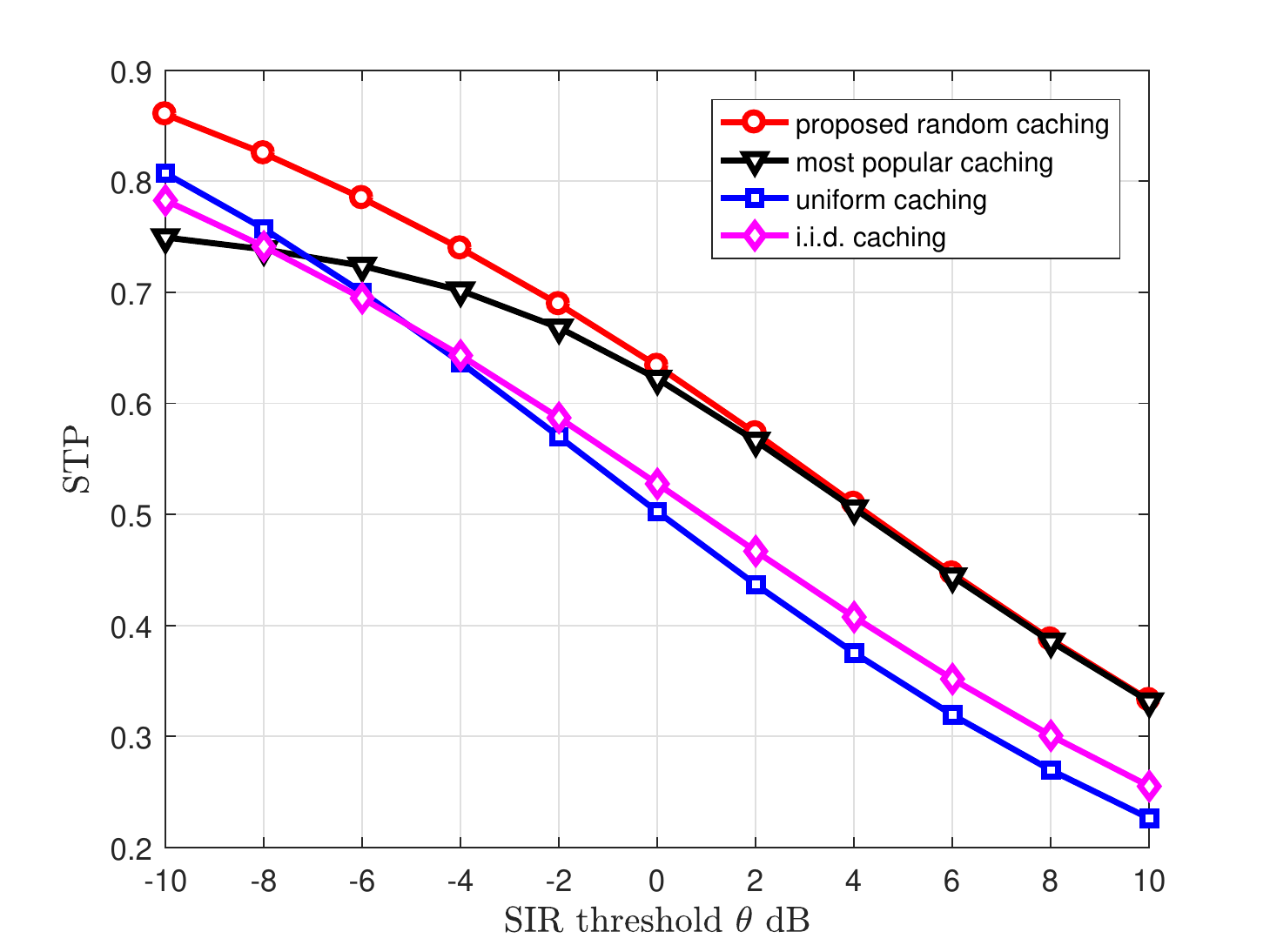}}

      \caption{STP versus $\theta$ at $\gamma=0.8$ and $M=5$.}\label{optSDPtheta}
\end{figure*}

In this part, we compare the {proposed solutions with three baselines in the two scenarios.}
Baseline 1 {adopts the most popular caching design where}  each BS in tier $k$ selects the $C_k$ most popular files to store \cite{Cache-enabledsmallcellnetworksmodelingandtradeoffs}.
Baseline 2 {adopts the uniform caching design, where} each BS in tier $k$ randomly selects $C_k$ files to store, according to the uniform distribution \cite{Tamoorulhassan2015Modeling}.
Baseline 3 {adopts the i.i.d. caching {design}, where}  each BS in tier $k$ randomly selects $C_k$ files to store, in an i.i.d. manner with file $n$ being selected with probability $a_n$ \cite{ALearningBasedApproachtoCachinginHeterogenousSmallCellNetworks}. {In addition, for each baseline}, in the high mobility scenario, the BS activity probability is {chosen as}  $\beta=1$ and in the static scenario, the  BS activity probability is obtained by  {solving Problem \ref{subProb2} for the {corresponding} caching probability $\mathbf{T}$ using the  method proposed in Section \ref{subsubsectionbetaST}.}

\begin{figure*}[!t]
    \centering
        \subfloat[High mobility scenario.]{\includegraphics[width=3.3in]{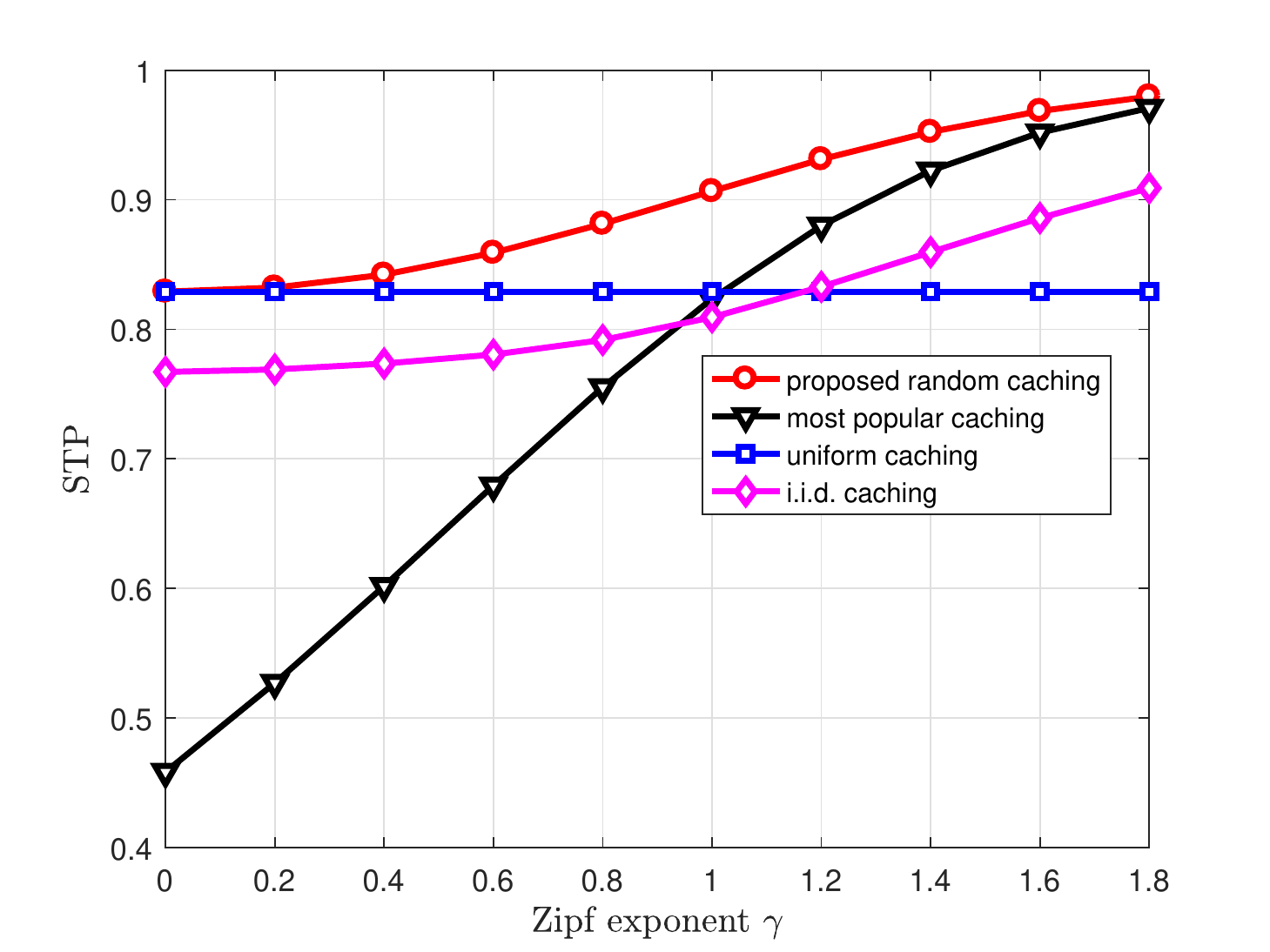}}
        \subfloat[Static scenario.]{\includegraphics[width=3.3in]{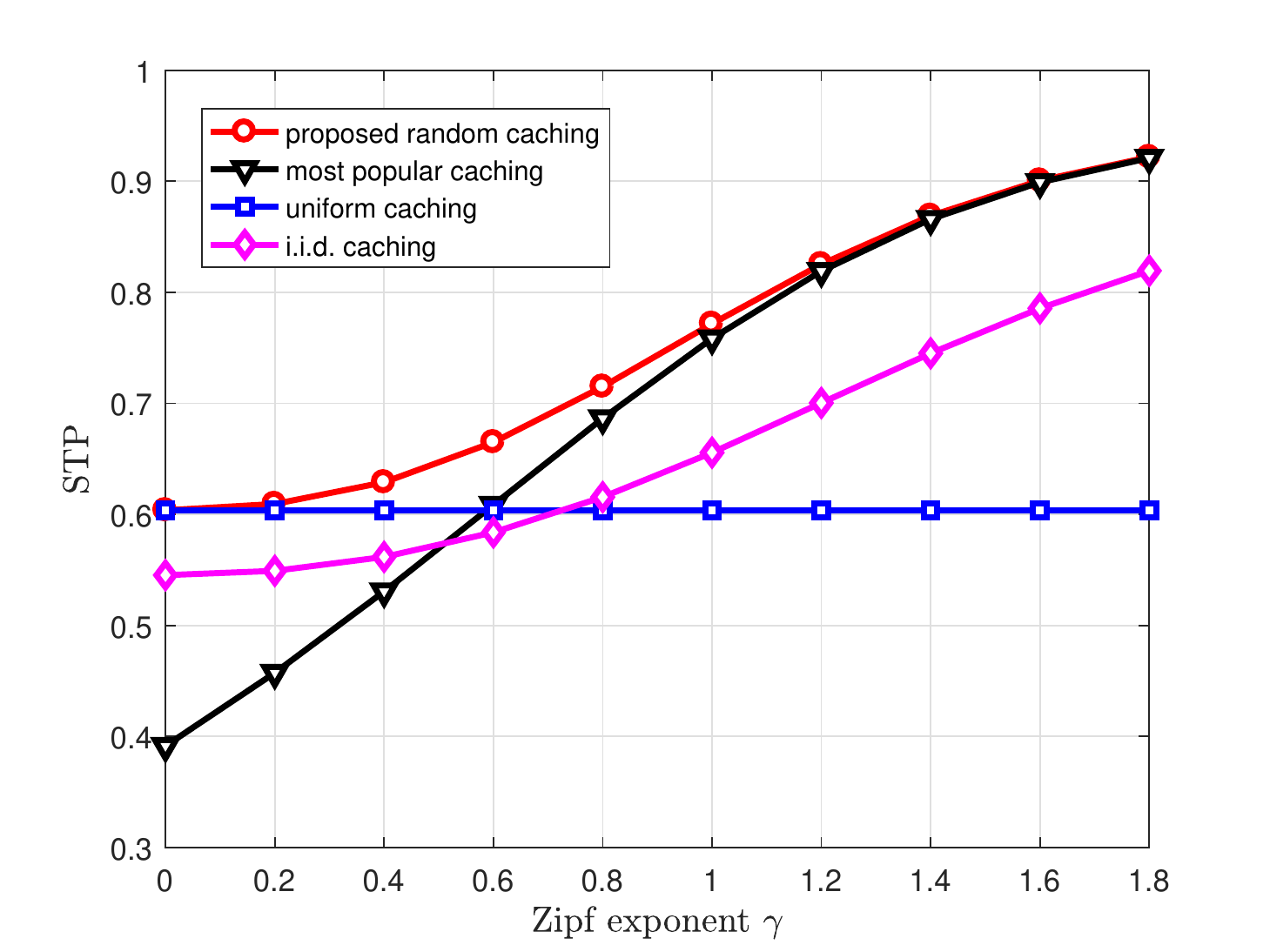}}
      \caption{STP versus $\gamma$ at $\theta=-3$ dB and $M=5$.}\label{optSDPgamma}
\end{figure*}

Fig. \ref{optSDPtheta} and Fig. \ref{optSDPgamma}  plot the STP versus the SIR threshold $\theta$ and the Zipf exponent $\gamma$, { respectively,  in the two} scenarios. Clearly, we see that the {proposed solutions} outperform all the baselines in both scenarios. In addition,  we can see that when  {$\theta$ is high or $\gamma$ is large}, the most popular caching can achieve almost the same performance as the proposed random caching; when  {$\theta$ is low or $\gamma$ is small}, the uniform caching can achieve almost the same performance as the proposed random caching. {The reasons are given as} follows. {The typical user can be successfully served by only the strongest BS when $\theta$ is high (the tail of the popularity distribution becomes small when $\gamma$ is large), and thus caching the most popular files at each BS is almost optimal; the typical user can be successfully served by more BSs when $\theta$ is low (the tail of the popularity distribution becomes flat when $\gamma$ is small), and thus caching more files in the network is better. } 



\section{Conclusions}
In this paper, we consider retransmissions with random DTX in a large-scale cache-enabled HetNet {employing random caching}.  We analyze
and optimize the STP in the high mobility and static scenarios, {and show that mobility increases temporal diversity, leading to the STP increase}. First, in each scenario,  we obtain closed-form expressions for the
STP in the general and low SIR threshold regimes. The analysis shows that a larger caching probability corresponds to a higher STP in both scenarios, which reveals the advantage of caching. In addition, the analysis reveals that random DTX can improve the STP in the static scenario and its benefit gradually diminishes when mobility increases.
Next, in each scenario, we consider the maximization of the STP with respect to the caching probability and the BS activity probability, which is a challenging non-convex optimization problem. We obtain a globally optimal solution in the high mobility scenario and a stationary point in the static scenario. Numerical results show that the proposed solutions achieve significant gains over existing baseline schemes and can well adapt to the changes of the system parameters to wisely utilize storage resources and transmission opportunities.     The practical situations are most likely to be somewhere between these two scenarios. The results here therefore have provided some theoretical performance bounds and insights for a practical large-scale cache-enabled  HetNet. More general practical scenarios will be explored in our future work.

\begin{appendices}
\section{Proof of Theorem \ref{TheoremSDPMb}}\label{ProofTheoremSDPMb}
For ease of illustration, we rewrite the SIR in (\ref{eqdefSIR}) as
{\begingroup\makeatletter\def\f@size{11}\check@mathfonts
\def\maketag@@@#1{\hbox{\m@th\normalsize\normalfont#1}}\setlength{\arraycolsep}{0.0em}
\begin{equation}\label{eqsir}
\mathrm{SIR}_{n,0}(t)=\frac{P_{k_0}h_{k_0,l_0,0}(t)X^{-\alpha}_{k_0,l_0,0}(t)\mathbbm{1}(l_0\in\mathcal{B}_{k_0}^a(t))} {I},
\end{equation}\setlength{\arraycolsep}{5pt}\endgroup}where $I\triangleq {I_{n,k_0}+I_{-n,k_0}+\sum\limits_{j=1,\neq k_0}^{K}\left(I_{n,j}+I_{-n,j}\right)}$ with $I_{n,k_0}\triangleq \sum_{l\in\Phi_{n,k_0}\setminus\{k_0\}}  P_{k_0}h_{k_0,l,0}(t) X_{k_0,l,0}^{-\alpha}(t) \\ \mathbbm{1}(l\in\mathcal{B}_{k_0}^a(t))$, $I_{-n,k_0}\triangleq \sum_{l\in\Phi_{-n,k_0}} P_{k_0}h_{k_0,l,0}(t)X_{k_0,l,0}^{-\alpha}(t) \mathbbm{1}(l\in\mathcal{B}_{k_0}^a(t))$, $I_{n,j}\triangleq \sum_{l\in\Phi_{n,j}}P_jh_{j,l,0}(t)\\X_{j,l,0}^{-\alpha}(t)\mathbbm{1}(l\in\mathcal{B}_j^a(t))$ and $I_{-n,j}\triangleq \sum_{l\in\Phi_{-n,j}}P_jh_{j,l,0}(t)X_{j,l,0}^{-\alpha}(t)\mathbbm{1}(l\in\mathcal{B}_j^a(t))$. Note that $\Phi_{-n,j}\triangleq \Phi_j\setminus \Phi_{n,j}$ for all $j\in\mathcal{K}$. Due to independent thinning induced by random caching, we know that $\Phi_{n,j}$ is a homogeneous PPP with density $\lambda_jT_{n,j}$ and $\Phi_{-n,j}$ is a homogeneous PPP with density $\lambda_j(1-T_{n,j})$. By (\ref{eqDefphihm}), we have
{\begingroup\makeatletter\def\f@size{11}\check@mathfonts
\def\maketag@@@#1{\hbox{\m@th\normalsize\normalfont#1}}\setlength{\arraycolsep}{0.0em}
\begin{eqnarray}
q_{n,\mathrm{hm}}(\mathbf{T}_n,\beta) &=& 1-\left(1-\Pr\left(\mathrm{SIR}_{n,0}\ge\theta\right)\right)^{M}.\label{eqEprhm}
\end{eqnarray}\setlength{\arraycolsep}{5pt}\endgroup}Thus, to calculate $q_{n,\mathrm{hm}}(\mathbf{T}_n,\beta)$, we only need to calculate $\Pr\left(\mathrm{SIR}_{n,0}\ge\theta\right)$.

First, we calculate $\Pr\left(\mathrm{SIR}_{n,0}\ge \theta|X_{k_0,l_0,0}=x\right)$. By (\ref{eqsir}), we have
{\begingroup\makeatletter\def\f@size{11}\check@mathfonts
\def\maketag@@@#1{\hbox{\m@th\normalsize\normalfont#1}}\setlength{\arraycolsep}{0.0em}
\begin{eqnarray}
&&\Pr\left(\mathrm{SIR}_{n,0}\ge \theta|X_{k_0,l_0,0}=x\right)\nonumber\\
&&\mathop=\limits^{(\mathrm{a})} \mathbbm{E}_{\{I_{n,k_0},I_{-n,k_0},I_{n,j},I_{-n,j}\}}\left(\exp\left(-\frac{\theta x^{\alpha}}{P_{k_0}}I \right)\right)\nonumber \\
&&\mathop=\limits^{(\mathrm{b})}  \underbrace{\mathbbm{E}_{I_{n,k_0}}\left(\exp\left(-\frac{\theta x^{\alpha}}{P_{k_0}}{I_{n,k_0}}\right)\right)}_{\triangleq \mathcal{L}_{I_{n,k_0}}(s)|_{s=\frac{\theta x^{\alpha}}{P_{k_0}}}} \underbrace{\mathbbm{E}_{I_{-n,k_0}}\left(\exp\left(-\frac{\theta x^{\alpha}}{P_{k_0}}{I_{-n,k_0}}\right)\right) }_{\triangleq \mathcal{L}_{I_{-n,k_0}}(s)|_{s=\frac{\theta x^{\alpha}}{P_{k_0}}}} \nonumber \\
&&{\times}\:\prod_{j=1,\neq k_0}^{K}\underbrace{\mathbbm{E}_{I_{n,j}}\left(\exp\left(-\frac{\theta x^{\alpha}}{P_{k_0}}{I_{n,j}}\right)\right) }_{\triangleq \mathcal{L}_{I_{n,j}}(s)|_{s=\frac{\theta x^{\alpha}}{P_{k_0}}}} \underbrace{\mathbbm{E}_{I_{-n,j}}\left(\exp\left(-\frac{\theta x^{\alpha}}{P_{k_0}}{I_{-n,j}}\right)\right)}_{\triangleq \mathcal{L}_{I_{-n,j}}(s)|_{s=\frac{\theta x^{\alpha}}{P_{k_0}}}}\label{eqPrsirfthm},
\end{eqnarray}\setlength{\arraycolsep}{5pt}\endgroup}where (a) is obtained by noting that $h_{k_0,l_0,0}$ is exponentially distributed
with unit mean; (b) is due to  the independence of the Rayleigh fading channels and the independence of the PPPs. $\mathcal{L}_{I_{n,k_0}}(s)$, $\mathcal{L}_{I_{-n,k_0}}(s)$, $\mathcal{L}_{I_{n,j}}(s)$ and $\mathcal{L}_{I_{-n,j}}(s)$ represent the Laplace transforms of the interference ${I_{n,k_0}}$,  ${I_{-n,k_0}}$, ${I_{n,j}}$ and ${I_{-n,j}}$, respectively, which are calculated as follows. 
{\begingroup\makeatletter\def\f@size{11}\check@mathfonts
\def\maketag@@@#1{\hbox{\m@th\normalsize\normalfont#1}}\setlength{\arraycolsep}{0.0em}
\begin{eqnarray}
\mathcal{L}_{I_{n,k_0}}(s)&=& \mathbbm{E}_{\Phi_{n,k_0}}\left(\prod_{l\in\Phi_{n,k_0}\setminus \{k_0\}} \mathbbm{E}_{h_{k_0,l,0}}\left(\exp\left(-sP_{k_0}h_{k_0,l,0} X^{-\alpha}_{k_0,l,0}\mathbbm{1}(l\in\mathcal{B}_j^a(t))\right)\right)\right)\nonumber \\
&\mathop=\limits^{(\mathrm{c})}&\exp\left(-\pi\lambda_{k_0}T_{n,k_0}x^2\beta\theta^{\frac{2}{\alpha}}\int_{\theta^{-\frac{2}{\alpha}}}^{\infty} \left(1-\frac{1}{1+v^{-\frac{\alpha}{2}}}\right)\mathrm{d}v\right)\nonumber\\
&=&\exp\left(-\pi\lambda_{k_0}T_{n,k_0}x^2\beta \left({_2}F_1\left(-\frac{2}{\alpha},1;1-\frac{2}{\alpha};-\theta\right)-1\right) \right) , \label{eqLIfthm}\end{eqnarray}\setlength{\arraycolsep}{5pt}\endgroup}where (c) is obtained by first noting that $h_{k_0,l_0,0}$ is exponentially distributed with unit mean and each interfering BS is active with probability $\beta$, and then utilizing the probability generating functional of PPP. Similarly, $\mathcal{L}_{I_{-n,k_0}}(s)$, $\mathcal{L}_{I_{n,j}}(s)$ and $\mathcal{L}_{I_{-n,j}}(s)$ can be calculated as follows.
{\begingroup\makeatletter\def\f@size{11}\check@mathfonts
\def\maketag@@@#1{\hbox{\m@th\normalsize\normalfont#1}}\setlength{\arraycolsep}{0.0em}
\begin{eqnarray}
\mathcal{L}_{I_{-n,k_0}}(s)&=&\exp\left(-\pi\lambda_{k_0}\left(1-T_{n,k_0}\right)x^2\beta \theta^{\frac{2}{\alpha}}\Gamma\left(1+\frac{2}{\alpha}\right)\Gamma\left(1-\frac{2}{\alpha}\right)\right),\label{eqLInonfthm2}\\
\mathcal{L}_{I_{n,j}}(s)&=&\exp\left(-\pi\lambda_jT_{n,j}\left(\frac{P_j}{P_{k_0}}\right)^{\frac{2}{\alpha}}x^2\beta \left({_2}F_1\left(-\frac{2}{\alpha},1;1-\frac{2}{\alpha};-\theta\right)-1\right)\right),\label{eqLInonfthm3} \\
\mathcal{L}_{I_{-n,j}}(s)&=&\exp\left(-\pi\lambda_j\left(1-T_{n,j}\right)\left(\frac{P_j}{P_{k_0}}\right)^{\frac{2}{\alpha}}x^2\beta \theta^{\frac{2}{\alpha}}\Gamma\left(1+\frac{2}{\alpha}\right)\Gamma\left(1-\frac{2}{\alpha}\right)\right).\label{eqLInonfthm4}
\end{eqnarray}\setlength{\arraycolsep}{5pt}\endgroup}Substituting (\ref{eqLIfthm})--(\ref{eqLInonfthm4}) into (\ref{eqPrsirfthm}), we can obtain $\Pr\left(\mathrm{SIR}_{n,0}\ge \theta|X_{k_0,l_0,0}=x\right)$.

Next, we calculate $\Pr\left(\mathrm{SIR}_{n,0}\ge \theta\right)$ by removing the condition $X_{k_0,l_0,0}=x$. Note that, we have the probability density function (p.d.f.) of the distance $x$, which is given by
{\begingroup\makeatletter\def\f@size{11}\check@mathfonts
\def\maketag@@@#1{\hbox{\m@th\normalsize\normalfont#1}}\setlength{\arraycolsep}{0.0em}
\begin{equation}\label{eqPDF}
f_{X_{k_0,l_0,0}}(x)=\frac{2\pi\lambda_{k_0}T_{n,k_0}}{A_{k_0}}x \exp\left(-\sum_{j=1}^{K}\pi\lambda_jT_{n,j}\left(\frac{P_j}{P_{k_0}}\right)^{\frac{2}{\alpha}}x^2\right),
\end{equation}\setlength{\arraycolsep}{5pt}\endgroup}where $A_{k_0}$ is the probability that the typical user $u_0$ is associated with tier $k_0$. Thus, we have
{\begingroup\makeatletter\def\f@size{11}\check@mathfonts
\def\maketag@@@#1{\hbox{\m@th\normalsize\normalfont#1}}\setlength{\arraycolsep}{0.0em}
\begin{eqnarray}
&&\Pr\left(\mathrm{SIR}_{n,0}\ge \theta \right)= \beta\int_{0}^{\infty}f_{X_{k_0,l_0,0}}(x)\left(\Pr\left(\mathrm{SIR}_{n,0}\ge \theta|X_{k_0,l_0,0}=x\right)\right)\mathrm{d}x\nonumber\\
&&\mathop=\limits^{(\mathrm{d})}\beta\sum_{k_0=1}^KA_{k_0}\int_{0}^{\infty}\exp\Bigg(-\sum_{j=1}^K \pi\lambda_j\left(\frac{P_j}{P_{k_0}}\right)^{\frac{2}{\alpha}}x ^2\bigg( T_{n,j}\beta \left({_2}F_1\left(-\frac{2}{\alpha},1;1-\frac{2}{\alpha};-\theta\right)-1\right) \nonumber\\
&&\quad+(1-T_{n,j})\beta \theta^{\frac{2}{\alpha}}\Gamma\left(1+\frac{2}{\alpha}\right)\Gamma\left(1-\frac{2}{\alpha}\right) \bigg)\Bigg) \frac{2\pi\lambda_{k_0}T_{n,k_0}}{A_{k_0}}x  \exp\left(-\sum_{j=1}^{K}\pi\lambda_jT_{n,j}\left(\frac{P_j}{P_{k_0}}\right)^{\frac{2}{\alpha}}x ^2\right)\mathrm{d}x \nonumber\\
&&\mathop=\limits^{(\mathrm{e})}\frac{\beta \sum_{k=1}^K\pi\lambda_kP_k^{\frac{2}{\alpha}}T_{n,k}}{\sum_{k=1}^{K}\pi\lambda_kP_k^{\frac{2}{\alpha}}\left(T_{n,k}W(\beta)+(1-T_{n,k})V(\beta)\right)}, \label{eqEprhmfinal}
\end{eqnarray}\setlength{\arraycolsep}{5pt}\endgroup}where (d) follows from the law of total probability and the serving BS is active with probability $\beta$; (e) follows from the definitions of $W(\beta)$ and $V(\beta)$ in Theorem \ref{TheoremSDPMb}.

Finally, substituting (\ref{eqEprhmfinal}) into (\ref{eqEprhm}), we  complete the proof of  Theorem~\ref{TheoremSDPMb}.

\section{Proof of Lemma \ref{lemmathetaarrow0mb}}\label{Prooflemmathetaarrow0hm}

In case i), $q_{n,\mathrm{hm}}(\mathbf{T}_n,\beta)$ in (\ref{eqCorollaryphiequalximb}) can be rewritten as
{\begingroup\makeatletter\def\f@size{11}\check@mathfonts
\def\maketag@@@#1{\hbox{\m@th\normalsize\normalfont#1}}\setlength{\arraycolsep}{0.0em}
\begin{equation*}\label{eqAPPqnhmrewcaseI}
q_{n,\mathrm{hm}}(\mathbf{T}_n,\beta)=1-\left(1-\frac{1}{{_2F}_1(-\frac{2}{\alpha},1;1-\frac{2}{\alpha};-\theta)}\right)^M.
\end{equation*}\setlength{\arraycolsep}{5pt}\endgroup}Then, when $\theta\to0$, we have
{\begingroup\makeatletter\def\f@size{11}\check@mathfonts
\def\maketag@@@#1{\hbox{\m@th\normalsize\normalfont#1}}\setlength{\arraycolsep}{0.0em}
\begin{eqnarray}\label{eqAPPcaseIinHM}
\bar{q}_{n,\mathrm{hm}}(\mathbf{T}_n,\beta)&\mathop\sim\limits^{(\mathrm{a})}&  \left(1-\frac{1}{1+\frac{2\theta}{ \alpha-2}}\right)^M \mathop\sim\limits^{(\mathrm{b})} \left(1-\left({1-\frac{2\theta}{\alpha-2}}\right)\right)^M=  \theta^M\left(\frac{2}{\alpha-2}\right)^M,
\end{eqnarray}\setlength{\arraycolsep}{5pt}\endgroup}where (a) is due to ${_2F}_1(-\frac{2}{\alpha},1;1-\frac{2}{\alpha};-\theta)\sim 1+\frac{2\theta}{\alpha-2}$ as $\theta\to0$ and (b) follows $\frac{1}{1+ax}\sim 1-ax$ as $x\to 0$.

In case ii), $q_{n,\mathrm{hm}}(\mathbf{T}_n,\beta)$ in (\ref{eqCorollaryphiequalximb}) can be rewritten as
{\begingroup\makeatletter\def\f@size{10}\check@mathfonts
\def\maketag@@@#1{\hbox{\m@th\normalsize\normalfont#1}}\setlength{\arraycolsep}{0.0em}
\begin{eqnarray*}\label{eqAPPrewcaseIIinHM}
&&q_{n,\mathrm{hm}}(\mathbf{T}_n,\beta)= 1-\left(1-\frac{\sum_{k\in\mathcal{K}}z_kT_{n,k}} {{_2F}_1(-\frac{2}{\alpha},1;1-\frac{2}{\alpha};-\theta) \sum\limits_{k\in\mathcal{K}}z_kT_{n,k} + \Gamma\left(1+\frac{2}{\alpha}\right)\Gamma\left(1-\frac{2}{\alpha}\right)\theta^{\frac{2}{\alpha}} (1-\sum\limits_{k\in\mathcal{K}}z_kT_{n,k})}\right)^M.
\end{eqnarray*}\setlength{\arraycolsep}{5pt}\endgroup}Then, when $\theta\to0$, we have
{\begingroup\makeatletter\def\f@size{11}\check@mathfonts
\def\maketag@@@#1{\hbox{\m@th\normalsize\normalfont#1}}\setlength{\arraycolsep}{0.0em}
\begin{eqnarray}\label{eqAPPcaseIIinHM}
\bar{q}_{n,\mathrm{hm}}(\mathbf{T}_n,\beta)&\mathop\sim\limits^{(\mathrm{c})}&  \left(1-\frac{\sum_{k\in\mathcal{K}}z_kT_{n,k}} {\sum_{k\in\mathcal{K}}z_kT_{n,k} \left(1+\frac{2\theta}{\alpha-2}\right) +\Gamma\left(1+\frac{2}{\alpha}\right)\Gamma\left(1-\frac{2}{\alpha}\right)\theta^{\frac{2}{\alpha}}(1-\sum_{k\in\mathcal{K}}z_kT_{n,k})}\right)^M\nonumber\\
&\mathop\sim\limits^{(\mathrm{d})}&  \left(1-\frac{1}{ 1+\left(\frac{1}{\sum_{k\in\mathcal{K}}z_kT_{n,k}}-1\right)\Gamma\left(1+\frac{2}{\alpha}\right)\Gamma\left(1-\frac{2}{\alpha}\right)\theta^{\frac{2}{\alpha}}}\right)^M\nonumber\\
&\mathop\sim\limits^{(\mathrm{e})}&  \theta^{\frac{2 M}{\alpha} }\left(\left(\frac{1}{\sum_{k\in\mathcal{K}}z_kT_{n,k}}-1\right)\Gamma\left(1+\frac{2}{\alpha}\right)\Gamma\left(1-\frac{2}{\alpha}\right)\right)^M,
\end{eqnarray}\setlength{\arraycolsep}{5pt}\endgroup}where (c) follows from ${{_2F}_1(-\frac{2}{\alpha},1;1-\frac{2}{\alpha};-\theta)}\sim 1+\frac{2\theta}{\alpha-2}$ as $\theta\to 0$; (d) uses the fact that the dominant term of the polynomial $c\theta+d\theta^{\frac{2}{\alpha}}$, $c,d>0$, $\alpha>2$, is $d\theta^{\frac{2}{\alpha}}$ when $\theta\to 0$; (e) is due to $\frac{1}{1+ax}\sim 1-ax$ as $x\to 0$.

In case iii), $q_{n,\mathrm{hm}}(\mathbf{T}_n,\beta)$ in (\ref{eqCorollaryphiequalximb}) can be rewritten as
{\begingroup\makeatletter\def\f@size{11}\check@mathfonts
\def\maketag@@@#1{\hbox{\m@th\normalsize\normalfont#1}}\setlength{\arraycolsep}{0.0em}
\begin{eqnarray*}\label{eqAPPrewcaseIIIinHM}
q_{n,\mathrm{hm}}(\mathbf{T}_n,\beta)=1-\left(1-\frac{\beta}{1-\beta+\beta{_2F}_1(-\frac{2}{\alpha},1;1-\frac{2}{\alpha};-\theta)}\right)^M.
\end{eqnarray*}\setlength{\arraycolsep}{5pt}\endgroup}Then, when $\theta\to0$, we have
{\begingroup\makeatletter\def\f@size{11}\check@mathfonts
\def\maketag@@@#1{\hbox{\m@th\normalsize\normalfont#1}}\setlength{\arraycolsep}{0.0em}
\begin{eqnarray}\label{eqAPPcaseIIIinHM}
\bar{q}_{n,\mathrm{hm}}(\mathbf{T}_n,\beta)&\sim&  \left(1-\frac{\beta}{1+\frac{2\theta\beta }{\alpha-2}}\right)^M\sim  \left(1-\beta\right)^M+\theta \left(1-\beta\right)^{M-1}M\beta^2\frac{2\theta }{\alpha-2},
\end{eqnarray}\setlength{\arraycolsep}{5pt}\endgroup}where the last step follows from the binomial expansion.

In case iv), based on $q_{n,\mathrm{hm}}(\mathbf{T}_n,\beta)$ in (\ref{eqCorollaryphiequalximb}), by using a similar method to case i) -- iii), when $\theta\to0$, we have
{\begingroup\makeatletter\def\f@size{10.5}\check@mathfonts
\def\maketag@@@#1{\hbox{\m@th\normalsize\normalfont#1}}\setlength{\arraycolsep}{0.0em}
\begin{eqnarray}\label{eqAPPcaseIVinHM}
\bar{q}_{n,\mathrm{hm}}(\mathbf{T}_n,\beta)\sim \left(1-\beta\right)^M+\theta^{\frac{2}{\alpha}}\left(1-\beta\right)^{M-1}M\beta^2\left(\frac{1}{\sum_{k\in\mathcal{K}}z_kT_{n,k}}-1\right) \Gamma\left(1+\frac{2}{\alpha}\right)\Gamma\left(1-\frac{2}{\alpha}\right).
\end{eqnarray}\setlength{\arraycolsep}{5pt}\endgroup}\vspace{-0.4cm}

Combining (\ref{eqAPPcaseIinHM})--(\ref{eqAPPcaseIVinHM}) and using the definition of  $c_{\mathrm{hm}}(\mathbf{T}_n,\beta)$ in (\ref{eqchm}), we complete the proof of Lemma \ref{lemmathetaarrow0mb}.

\section{Proof of Theorem \ref{TheoremSDP}}\label{ProofTheoremSDP}
By using the binomial expansion, the STP in (\ref{eqDefphist}) can be rewritten as
{\begingroup\makeatletter\def\f@size{11}\check@mathfonts
\def\maketag@@@#1{\hbox{\m@th\normalsize\normalfont#1}}\setlength{\arraycolsep}{0.0em}
\begin{eqnarray}\label{eqphifstinapp}
q_{n,\mathrm{st}}(\mathbf{T}_n,\beta)&=&\sum_{m=1}^{M}\binom{M}{m} (-1)^{m+1} \mathbbm{E}_{\Phi}\left(\left(\Pr\left(\mathrm{SIR}_{n,0}\ge \theta|\Phi\right)\right)^m\right),
\end{eqnarray}\setlength{\arraycolsep}{5pt}\endgroup}where $\mathrm{SIR}_{n,0}$ is given by (\ref{eqsir}). Thus, to calculate $q_{n,\mathrm{st}}(\mathbf{T}_n,\beta)$ in (\ref{eqphifstinapp}), it remains to calculate $\mathbbm{E}_{\Phi}\left(\left(\Pr\left(\mathrm{SIR}_{n,0}\ge \theta|\Phi\right)\right)^m\right)$. As in (\ref{eqPrsirfthm}), conditioning on
$X_{k_0,l_0,0}=x$, we have
{\begingroup\makeatletter\def\f@size{11}\check@mathfonts
\def\maketag@@@#1{\hbox{\m@th\normalsize\normalfont#1}}\setlength{\arraycolsep}{0.0em}
\begin{eqnarray}\label{eqEphi}
&&\mathbbm{E}_{\Phi}\left(\left(\Pr\left(\mathrm{SIR}_{n,0}\ge \theta|\Phi,X_{k_0,l_0,0}=x\right)\right)^m\right)\nonumber\\
&&= \mathbbm{E}_{\Phi}\left(\left(\mathbbm{E}_{\{I_{n,k_0|\Phi},I_{-n,k_0|\Phi},I_{n,j|\Phi},I_{-n,j|\Phi}\}}\left(\exp\left(-\frac{\theta x^{\alpha}}{P_{k_0}}I \right)\right)\right)^m\right)\nonumber \\
&&= \mathbbm{E}_{\Phi}\Bigg(\bigg(\underbrace{\mathbbm{E}_{I_{n,k_0|\Phi}}\left(\exp\left(-\frac{\theta x^{\alpha}}{P_{k_0}}I_{n,k_0|\Phi} \right)\right)}_{\triangleq \mathcal{L}_{I_{n,k_0|\Phi}}(s)|_{s=\frac{\theta x^{\alpha}}{P_{k_0}}}} \underbrace{\mathbbm{E}_{I_{-n,k_0|\Phi}}\left(\exp\left(-\frac{\theta x^{\alpha}}{P_{k_0}}I_{-n,k_0|\Phi} \right)\right)}_{\triangleq \mathcal{L}_{I_{-n,k_0|\Phi}}(s)|_{s=\frac{\theta x^{\alpha}}{P_{k_0}}}} \nonumber\\
&&{\times}\:\prod_{j=1,\neq k_0}^{K}\underbrace{\mathbbm{E}_{ I_{n,j|\Phi}}\left(\exp\left(-\frac{\theta x^{\alpha}}{P_{k_0}}I_{n,j|\Phi} \right)\right) }_{\triangleq \mathcal{L}_{I_{n,j|\Phi}}(s)|_{s=\frac{\theta x^{\alpha}}{P_{k_0}}}} \underbrace{\mathbbm{E}_{I_{-n,j|\Phi}}\left(\exp\left(-\frac{\theta x^{\alpha}}{P_{k_0}}I_{-n,j|\Phi} \right)\right)}_{\triangleq \mathcal{L}_{I_{-n,j|\Phi}}(s)|_{s=\frac{\theta x^{\alpha}}{P_{k_0}}}}\bigg)^m\Bigg),
\end{eqnarray}\setlength{\arraycolsep}{5pt}\endgroup}where  $\mathcal{L}_{I_{n,k_0|\Phi}}(s)$, $\mathcal{L}_{I_{-n,k_0|\Phi}}(s)$, $\mathcal{L}_{I_{n,j|\Phi}}(s)$ and $\mathcal{L}_{I_{-n,j|\Phi}}(s)$ represent the Laplace transforms of the interference ${I_{n,k_0|\Phi}}$,  ${I_{-n,k_0|\Phi}}$, ${I_{n,j|\Phi}}$ and ${I_{-n,j|\Phi}}$, conditioned on $\Phi$, respectively, which can be calculated as 
{\begingroup\makeatletter\def\f@size{10}\check@mathfonts
\def\maketag@@@#1{\hbox{\m@th\normalsize\normalfont#1}}\setlength{\arraycolsep}{0.0em}
\begin{eqnarray*}
\mathcal{L}_{I_{n,k_0|\Phi}}(s)&=& \prod_{l\in\Phi_{n,k_0}\setminus\{l_0\}} \left(\frac{\beta}{1+sP_j X^{-\alpha}_{k_0,l,0}}+1-\beta\right),
\mathcal{L}_{I_{-n,k_0|\Phi}}(s)= \prod_{l\in\Phi_{-n,k_0}} \left(\frac{\beta}{1+sP_j X^{-\alpha}_{k_0,l,0}}+1-\beta\right),\label{eqLIft1}\\
\mathcal{L}_{I_{n,j|\Phi}}(s)&=& \prod_{l\in\Phi_{n,j}} \left(\frac{\beta}{1+sP_j X^{-\alpha}_{j,l,0}}+1-\beta\right),
\mathcal{L}_{I_{-n,j|\Phi}}(s)=\prod_{l\in\Phi_{-n,j}} \left(\frac{\beta}{1+sP_j X^{-\alpha}_{j,l,0}}+1-\beta\right).\label{eqLIft4}
\end{eqnarray*}\setlength{\arraycolsep}{5pt}\endgroup}

Next, we calculate $\mathbbm{E}_{\Phi}\left(\left(\Pr\left(\mathrm{SIR}_{n,0}\ge \theta|\Phi\right)\right)^m\right)$ by removing the condition $X_{k_0,l_0,0}=x$ in $\mathbbm{E}_{\Phi}\left(\left(\Pr\left(\mathrm{SIR}_{n,0}\ge \theta|\Phi,X_{k_0,l_0,0}=x\right)\right)^m\right)$. By using the p.d.f. of  $X_{k_0,l_0,0}$ in (\ref{eqPDF}) and applying some similar algebraic manipulations as used in  (\ref{eqEprhmfinal}), we have
{\begingroup\makeatletter\def\f@size{11}\check@mathfonts
\def\maketag@@@#1{\hbox{\m@th\normalsize\normalfont#1}}\setlength{\arraycolsep}{0.0em}
\begin{eqnarray}\label{eqEpr2}
&&\mathbbm{E}_{\Phi}\left(\left(\Pr\left(\mathrm{SIR}_{n,0}\ge \theta|\Phi\right)\right)^m\right)=\beta^m\int_{0}^{\infty} f_{X_{k_0,l_0,0}}(x) \mathbbm{E}_{\Phi}\left(\left(\Pr\left(\mathrm{SIR}_{n,0}\ge \theta|\Phi,X_{k_0,l_0,0}=x\right)\right)^m\right)\mathrm{d}x\nonumber\\
&&\mathop=\limits^{(\mathrm{a})}{\beta^m \sum_{k=1}^K\pi\lambda_kT_{n,k}} \Bigg(\sum_{j=1}^K\pi\lambda_j\left(\frac{P_j}{P_k}\right)^{\frac{2}{\alpha}}\bigg(T_{n,j}\theta^{\frac{2}{\alpha}}\int_{\theta^{-\frac{2}{\alpha}}}^{\infty} \left(1-\left(\frac{\beta}{1+v^{-\frac{\alpha}{2}}}+1-\beta\right)^m\right)\mathrm{d}v\nonumber\\
&&\quad+ (1-T_{n,j})\theta^{\frac{2}{\alpha}}\int_{0}^{\infty}\left(1-\left(\frac{\beta}{1+v^{-\frac{\alpha}{2}}}+1-\beta\right)^m\right)\mathrm{d}v+T_{n,j}\bigg) \Bigg)^{-1}\nonumber\\
&&\mathop=\limits^{(\mathrm{b})}\frac{\beta^m \sum_{k=1}^K\pi\lambda_kP_k^{\frac{2}{\alpha}}T_{n,k}}{\sum_{k=1}^{K}\pi\lambda_kP_k^{\frac{2}{\alpha}}\left(T_{n,k}F_m(\beta)+(1-T_{n,k})G_m(\beta)\right)},
\end{eqnarray}\setlength{\arraycolsep}{5pt}\endgroup}where (a) is obtained by  utilizing the probability generating functional of PPP; (b) follows from
{\begingroup\makeatletter\def\f@size{11}\check@mathfonts
\def\maketag@@@#1{\hbox{\m@th\normalsize\normalfont#1}}\setlength{\arraycolsep}{0.0em}
\begin{eqnarray*}
\int_{\theta^{-\frac{2}{\alpha}}}^{\infty} \left(1-\left(\frac{\beta}{1+v^{-\frac{\alpha}{2}}}+1-\beta\right)^m\right)\mathrm{d}v&=& \sum_{i=0}^{m}\binom{m}{i}\beta^i(1-\beta)^{m-i}\left({_2}F_1\left(-\frac{2}{\alpha},i;1-\frac{2}{\alpha};-\theta\right)-1\right)\theta^{-\frac{2}{\alpha}},\nonumber\\
\int_0^{\infty} \left(1-\left(\frac{\beta}{1+v^{-\frac{\alpha}{2}}}+1-\beta\right)^m\right)\mathrm{d}v&=& \sum_{i=0}^{m}\binom{m}{i}\beta^i(1-\beta)^{m-i}\frac{\Gamma\left(i+\frac{2}{\alpha}\right)}{\Gamma(i)}\Gamma\left(1-\frac{2}{\alpha}\right),
\end{eqnarray*}\setlength{\arraycolsep}{5pt}\endgroup}and the definitions of $F_m(\beta)$ and $G_m(\beta)$ in (\ref{eqFm}) and (\ref{eqGm}).

Finally, substituting (\ref{eqEpr2}) into (\ref{eqphifstinapp}), we complete the proof of Theorem \ref{TheoremSDP}.

\section{Proof of Lemma \ref{lemmathetaarrow0}}\label{Prooflemmathetaarrow0}
In case i), $q_{n,\mathrm{st}}(\mathbf{T}_n,\beta)$ in (\ref{eqCorollaryphiequalxi}) can be rewritten as
{\begingroup\makeatletter\def\f@size{11}\check@mathfonts
\def\maketag@@@#1{\hbox{\m@th\normalsize\normalfont#1}}\setlength{\arraycolsep}{0.0em}
\begin{eqnarray*}\label{eqphifstappre2}
q_{n,\mathrm{st}}(\mathbf{T}_n,\beta)=\sum_{m=1}^{M}\binom{M}{m}(-1)^{m+1}\frac{1}{{_2}F_1(-\frac{2}{\alpha},m;1-\frac{2}{\alpha};-\theta)}.
\end{eqnarray*}\setlength{\arraycolsep}{5pt}\endgroup}Then, by following the similar steps to those for proving Proposition 3 in \cite{AStochasticGeometryAnalysisofInterCellInterferenceCoordinationandIntraCellDiversity}, when $\theta\rightarrow0$, we have
{\begingroup\makeatletter\def\f@size{11}\check@mathfonts
\def\maketag@@@#1{\hbox{\m@th\normalsize\normalfont#1}}\setlength{\arraycolsep}{0.0em}
\begin{eqnarray}\label{eqphifstTfis1case2}
\bar{q}_{n,\mathrm{st}}(\mathbf{T}_n,\beta)&\sim& \theta^M\frac{\partial^M}{\partial x^M}\left({_1F_1}(-\frac{2}{\alpha};1-\frac{2}{\alpha};x)\right)^{-1}|_{x=0}.
\end{eqnarray}\setlength{\arraycolsep}{5pt}\endgroup}We have omitted the details due to page limitation.

In cases ii) and iv), based on (\ref{eqCorollaryphiequalxi}), when $\theta\to0$, we have
{\begingroup\makeatletter\def\f@size{11}\check@mathfonts
\def\maketag@@@#1{\hbox{\m@th\normalsize\normalfont#1}}\setlength{\arraycolsep}{0.0em}
\begin{eqnarray}\label{eqphifstTfnot1}
\bar{q}_{n,\mathrm{st}}(\mathbf{T}_n,\beta)&\mathop\sim\limits^{(\mathrm{a})}& 1-\sum_{m=1}^{M}\binom{M}{m}\frac{(-1)^{m+1}\beta^m}{ \sum_{i=0}^{m}\binom{m}{i}\beta^i\left(1-\beta\right)^{m-i}\left(1+(\frac{1}{\sum_{k\in\mathcal{K}}z_kT_{n,k}}-1) \frac{\Gamma\left(i+\frac{2}{\alpha}\right)}{\Gamma(i)}\Gamma\left(1-\frac{2}{\alpha}\right)\theta^{\frac{2}{\alpha}}\right)}\nonumber\\
&\mathop\sim\limits^{(\mathrm{b})}& 1-\sum_{m=1}^{M}\binom{M}{m}{(-1)^{m+1}\beta^m}\Bigg(1- \left(\frac{1}{\sum_{k\in\mathcal{K}}z_kT_{n,k}}-1\right)\Gamma\left(1-\frac{2}{\alpha}\right)\nonumber\\
&&{\times}\;\theta^{\frac{2}{\alpha}} \sum_{i=0}^{m}\binom{m}{i}\beta^i\left(1-\beta\right)^{m-i}\frac{\Gamma\left(i+\frac{2}{\alpha}\right)}{\Gamma(i)}\Bigg)\nonumber\\
&\mathop=\limits^{(\mathrm{c})}&\left(1-\beta\right)^M+ \theta^{\frac{2}{\alpha}}\left(\frac{1}{\sum_{k\in\mathcal{K}}z_kT_{n,k}}-1\right) \Gamma\left(1-\frac{2}{\alpha}\right)\nonumber\\
&&{\times}\;\sum_{m=1}^{M}\binom{M}{m}{(-1)^{m+1}\beta^m} \sum_{i=0}^{m}\binom{m}{i}\beta^i\left(1-\beta\right)^{m-i}\frac{\Gamma\left(i+\frac{2}{\alpha}\right)}{\Gamma(i)},
\end{eqnarray}\setlength{\arraycolsep}{5pt}\endgroup}where (a) follows from ${_2}F_1(-\frac{2}{\alpha},i;1-\frac{2}{\alpha};-\theta)\sim 1+\frac{2i\theta}{\alpha-2}$ when $\theta\to 0$ and  is based on the fact that the dominant term of the polynomial $c\theta+d\theta^{\frac{2}{\alpha}}$, $c,d>0$, $\alpha>2$, is $d\theta^{\frac{2}{\alpha}}$ when $\theta\to 0$; (b) is due to the fact that $\sum_{i=0}^{m}\binom{m}{i}\beta^i\left(1-\beta\right)^{m-i}=1$ and $\frac{1}{1+ax}\sim 1-ax$ when $x\rightarrow 0$; (c) follows from $\sum_{m=1}^{M}\binom{M}{m}(-1)^{m+1}\beta^m=1-(1-\beta)^M$.

In case iii), $q_{n,\mathrm{st}}(\mathbf{T}_n,\beta)$ in (\ref{eqCorollaryphiequalxi}) can be rewritten as
{\begingroup\makeatletter\def\f@size{11}\check@mathfonts
\def\maketag@@@#1{\hbox{\m@th\normalsize\normalfont#1}}\setlength{\arraycolsep}{0.0em}
\begin{eqnarray*}\label{eqphifstappre}
q_{n,\mathrm{st}}(\mathbf{T}_n,\beta)=\sum_{m=1}^{M}\binom{M}{m}(-1)^{m+1}\frac{\beta^m}{F_m(\beta)}.
\end{eqnarray*}\setlength{\arraycolsep}{5pt}\endgroup}Then, when $\theta\to0$, we easily have
{\begingroup\makeatletter\def\f@size{11}\check@mathfonts
\def\maketag@@@#1{\hbox{\m@th\normalsize\normalfont#1}}\setlength{\arraycolsep}{0.0em}
\begin{eqnarray}\label{eqphifstTfis1casei}
\bar{q}_{n,\mathrm{st}}(\mathbf{T}_n,\beta)
&\mathop\sim\limits^{(\mathrm{d})}& 1-\sum_{m=1}^{M}\binom{M}{m}\frac{(-1)^{m+1}\beta^m}{ 1+\frac{2\theta}{\alpha-2}m\beta } \sim 1-\sum_{m=1}^{M}\binom{M}{m}{(-1)^{m+1}\beta^m}\left( 1-\frac{2\theta}{\alpha-2}m\beta \right)\nonumber\\
&\mathop=\limits^{(\mathrm{e})}&  \left(1-\beta\right)^M+\theta(1-\beta)^{M-1}M\beta^2\frac{2}{\alpha-2},
\end{eqnarray}\setlength{\arraycolsep}{5pt}\endgroup}where (d) follows from ${_2}F_1(-\frac{2}{\alpha},i;1-\frac{2}{\alpha};-\theta)\sim 1+\frac{2i\theta}{\alpha-2}$ as $\theta\rightarrow0$ and is based on the fact that $\sum_{i=0}^{m}\binom{m}{i}\beta^i\left(1-\beta\right)^{m-i}i=m\beta$; (e) follows from $\sum_{m=1}^{M}\binom{M}{m}(-1)^{m+1}m\beta^{m+1}=(1-\beta)^{M-1}M\beta^2$.

Combining (\ref{eqphifstTfis1case2})--(\ref{eqphifstTfis1casei}) and using the definition of $c_{\mathrm{st}}(\mathbf{T}_n,\beta)$ in (\ref{eqcst}), we complete the proof of Lemma~\ref{lemmathetaarrow0}.

\end{appendices}

%

\end{document}